\documentclass[aps,prl,reprint,superscriptaddress,floatfix]{revtex4-2}
\usepackage[colorlinks=true,allcolors=blue]{hyperref} 
\usepackage{graphicx} 
\usepackage{bm}% bold math
\usepackage{amsmath, amssymb}
\usepackage{mathtools}
\usepackage{braket}
\usepackage{appendix}
\usepackage{siunitx}

\newcommand{\Figref}[1]{Fig.~\ref{#1}}

\newcommand{\Eqref}[1]{Eq.~(\ref{#1})}

\newcommand{\lb}{\left(}
\newcommand{\rb}{\right)}
\newcommand{\LB}{\left[}
\newcommand{\RB}{\right]}

\newcommand{\added}[1]{{\color{black}{#1}}}

\newcommand{\tk}[1]{{\color{black}{#1}}}
\newcommand{\ytm}[1]{{\color{black}{#1}}}
\newcommand{\kb}[1]{{\color{black}{#1}}}

\begin{document}
\title{Directional Symmetry Breaking of Spherical Active Colloids by Magnetoviscous Coupling}
%\title{Hydrodynamics of Janus Particles in Ferrofluids}
\author{Ziyang Zhou}
\affiliation{Department of Chemical Science and Engineering, Kyoto University, Kyoto 615-8510, Japan}

\author{Takuya Kobayashi}
\affiliation{Department of Chemical Science and Engineering, Kyoto University, Kyoto 615-8510, Japan}
\affiliation{Department of Chemical Engineering, Stanford University, Stanford, CA 94305, USA}

\author{Keita Saito}
\affiliation{RIKEN Center for Emergent Matter Science, 2-1 Hirosawa, Wako, Saitama 351-0198, Japan}

\author{Masato Ito}
\affiliation{Department of Chemical Science and Engineering, Kyoto University, Kyoto 615-8510, Japan}

\author{Kohei Yoshinaga}
\affiliation{Department of Chemistry, Graduate School of Science, Kyoto University, Kyoto 606-8502, Japan}

\author{Yasutaka Iwashita}
\affiliation{Department of Physics, School of Science, Kyoto Sangyo University, Kyoto 603-8555, Japan}

\author{Kazusa Beppu}\email{kazusa.beppu@cheme.kyoto-u.ac.jp}
\affiliation{Department of Chemical Science and Engineering, Kyoto University, Kyoto 615-8510, Japan}

\author{Yusuke T. Maeda}
\email{maeda@cheme.kyoto-u.ac.jp}
\affiliation{Department of Chemical Science and Engineering, Kyoto University, Kyoto 615-8510, Japan}

\date{\today}

\begin{abstract}
Harnessing active matter calls for strategies that break the directional symmetry of self-propelled motion without altering the propulsion mechanism itself. Here, we show that magnetically inert spherical active colloids can be steered through the anisotropic viscous response of a ferrofluid under a uniform magnetic field. Self-propelled Janus colloids exhibit robust cross-field motion transverse to the magnetic field, although the applied magnetic field directly controls neither the particles nor their propulsion speed. Quantitative measurements reveal an emergent reorientation torque that grows with both propulsion speed and magnetic field strength. A squirmer model in a magnetoviscous medium captures these observations and shows that the torque arises from the coupling between swimmer-generated flow and anisotropic rotational viscosity. Our findings establish a hydrodynamic basis for converting viscous dissipation into directional symmetry breaking through anisotropic rheology, providing a route to field-controlled material transport by active matter.
\end{abstract}

\maketitle

A central challenge in active matter is to understand how self-propelled motion is generated and reoriented, because self-propulsion not only governs the collective states of active matter~\cite{marchetti2013hydrodynamics,michelin2023self,elgeti2015physics,spagnolie2023swimming} but also underlies transport~\cite{boymelgreen2018active,demirors2018active,ussia2024magnetic,song2025magnetically} and delivery~\cite{ghosh2020active,gao2014synthetic} at the microscale. Self-propelled colloidal particles (active colloids) provide a useful class of active matter, as their motion and interactions can be systematically controlled~\cite{Palacci2013Living,simmchen2016topographical,bechinger2016active,liebchen2018synthetic,katuri2018cross,Linden2019Interrupted,Auschra2021Thermotaxis,Zhang2021}. Yet the physical principles that enable robust directional control, particularly in \added{anisotropic fluid} media~\cite{lavrentovich2016active,aranson2018harnessing,bukusoglu2016design}, remain less understood than those governing self-propulsion.

\added{An open question is whether anisotropic viscous response alone can reorient an active particle and thereby break the directional symmetry of its motion. In theory, the flow field generated by self-propulsion can couple to anisotropic rheology, giving rise to a nontrivial torque even on a spherical swimmer, as predicted for a uniaxially aligned liquid-crystalline medium~\cite{Lintuvuori2017-hq,Daddi2018Dynamics}. However, in experiments on conventional liquid-crystalline media, the contribution of this mechanism has not yet been disentangled from other effects, because swimmer motion is influenced not only by anisotropic viscosity associated with orientational order~\cite{zhou2014living,guillamat2016control,zhou2017dynamic,genkin2017topological,turiv2020polar,koizumi2020control,ma2021programmable} but also by topological defects~\cite{peng2016command,rajabi2021directional,lesniewska2022controllable,sahu2022electrophoresis}, viscoelasticity~\cite{smalyukh2008elasticity,chi2022interaction,baza2026bend}, anchoring~\cite{chi2020surface,sahu2020omnidirectional,Sudha2024Behavior}, and confinement~\cite{mandal2025cooperativity}.}

\added{An attractive route to this problem is provided by ferrofluids, which are suspensions of magnetic nanoparticles (MNPs)~\cite{shliomis1974magnetic,Odenbach2000magnetoviscous,odenbach2009colloidal,Mahle2008-hw}. Under a uniform magnetic field, they acquire a tunable and reversible uniaxial viscous anisotropy through the field-induced response of magnetic nanoparticles~\cite{cherian2021electroferrofluids,socoliuc2022ferrofluids,wang2021magnetic,beppu2024magnetically}. With this anisotropy imposed externally and varied systematically, ferrofluids provide a well-controlled framework for probing how self-generated flows couple to anisotropic viscosity. In addition, the energy input for particle propulsion can be supplied independently of the field that controls the medium rheology, allowing flow-induced reorientation and directional symmetry breaking in anisotropic media to be identified more directly.}

In this Letter, we report robust cross-field motion, transverse to the applied uniform magnetic field, of magnetically inert spherical active colloids dispersed in ferrofluids. We show that this reorientation arises from an emergent torque that appears only when the particles are self-propelled and the ferrofluid exhibits a uniaxial viscous anisotropy. A squirmer model, combined with experiment, reveals a general mechanism of directional symmetry breaking in which swimmer-generated flow couples to anisotropic rotational viscosity to produce a reorientation torque.

\begin{figure}[tb]
\begin{center}
\includegraphics[width=8.5cm]{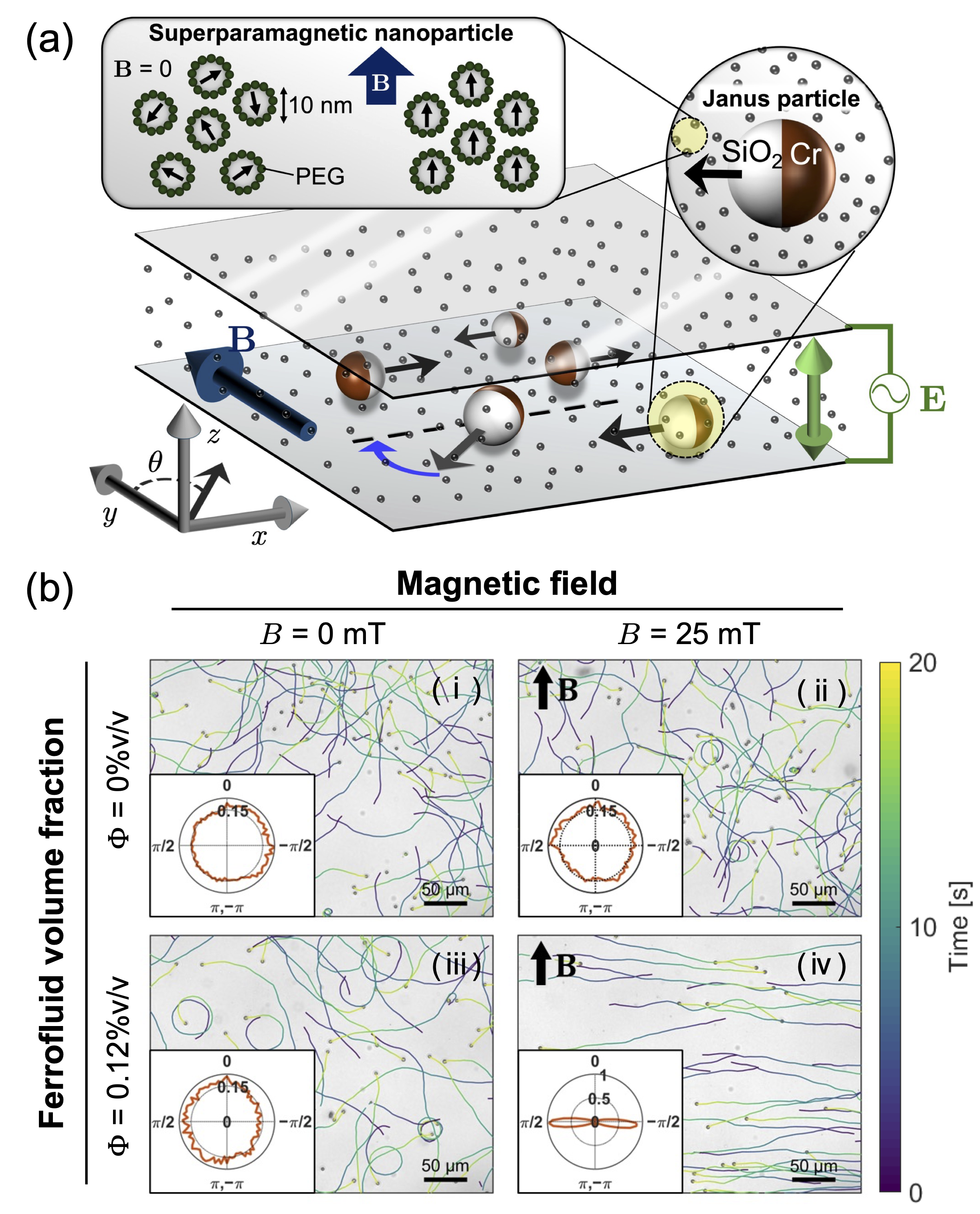}
\caption{(a) The experimental setup. The magnetic field is applied along the $y$ direction ($\theta = 0$). (b) Magnetic-field and ferrofluid dependence of the swimming direction. Trajectories of Janus colloids are color-coded. Inset: Probability distribution of the swimming direction $\theta_v$. Cross-field motion ($\theta_v=\pm \pi/2$) is observed only in the ferrofluid with $\Phi=0.12$\%v/v under an applied magnetic field, corresponding to condition ($\mathrm{iv}$).}
\label{fig1}
\end{center}
\end{figure}

We used self-propelled Janus colloids with a chromium (Cr)-coated hemisphere, with a diameter of \SI{3.6}{\micro\meter}, dispersed at a density of 0.67$\pm$0.15 particles/\SI{1000}{\micro\meter}$^2$. Self-propulsion via induced charge electrophoresis (ICEP)~\cite{bazant2004induced,squires2004induced,Gangwal2008-db} was driven by an AC electric field $E$ with a fixed frequency of 2 kHz applied along the $z$ direction through ITO glass electrodes (Fig.~\ref{fig1}(a)). The Janus colloids were dispersed either in an aqueous solution or in a dilute ferrofluid (PBG300, Ferrotec) containing sterically stabilized iron-oxide MNPs at a volume fraction of $\Phi=0.12\%$v/v. We applied a uniform magnetic field along the $y$ axis using a pair of horizontally aligned electromagnets, taking the magnetic-field direction as $\theta=0$. The magnetic flux density is $\mathbf{B}=B\bm{e}_y$, where $\bm{e}_y$ denotes the unit vector along the $y$ direction. The swimming direction of each particle is denoted by $\theta_v$, while the polarity angle $\theta_p$ is defined by image analysis of the polarity axis set by the Cr-coated hemisphere. Both $\theta_v$ and $\theta_p$ are measured from the magnetic-field direction. We confirmed that $\theta_p$ closely agrees with $\theta_v$ (Fig.~S1).

Fig.~\ref{fig1}(b) establishes the minimal conditions required for field-controlled reorientation of active colloids. In an aqueous solution, the swimming direction $\theta_v$ is isotropically distributed both without and with an applied magnetic field (Fig.~\ref{fig1}(b)-($\mathrm{i}$) and ($\mathrm{ii}$), respectively, Movie~1), showing that the Janus colloids themselves are magnetically inert under the present conditions. In ferrofluid, the directional distribution likewise remains isotropic in the absence of a magnetic field (Fig.~\ref{fig1}(b)-($\mathrm{iii}$), Movie~1), indicating that MNPs alone do not bias the swimming direction. Cross-field motion emerges only when self-propelled particles swim in a ferrofluid under an applied magnetic field (Fig.~\ref{fig1}(b)-($\mathrm{iv}$), Movie~1), where the swimming direction becomes strongly biased toward $\theta_v=\pm\pi/2$. 

In the absence of AC driving, the polarity angle $\theta_p$ remains isotropically distributed even in the ferrofluid under $B=\SI{25}{\milli\tesla}$ (Fig.~S2). Thus, in the absence of self-generated flows, we find no detectable field-induced orientational bias of the Janus colloids under the present conditions.
\ytm{The propulsion speed is unchanged by either the magnetic field or the addition of MNPs (Figs.~S3 and S4).} The same cross-field motion is also observed for Au-coated Janus colloids (Fig.~S5), further indicating that the effect does not rely on the weak magnetic susceptibility of the Cr coating. The reorientation therefore requires both the ferrofluid and the applied magnetic field, and does not occur when either of them is absent.

\begin{figure*}[tb]
\begin{center}
\includegraphics[width=18cm]{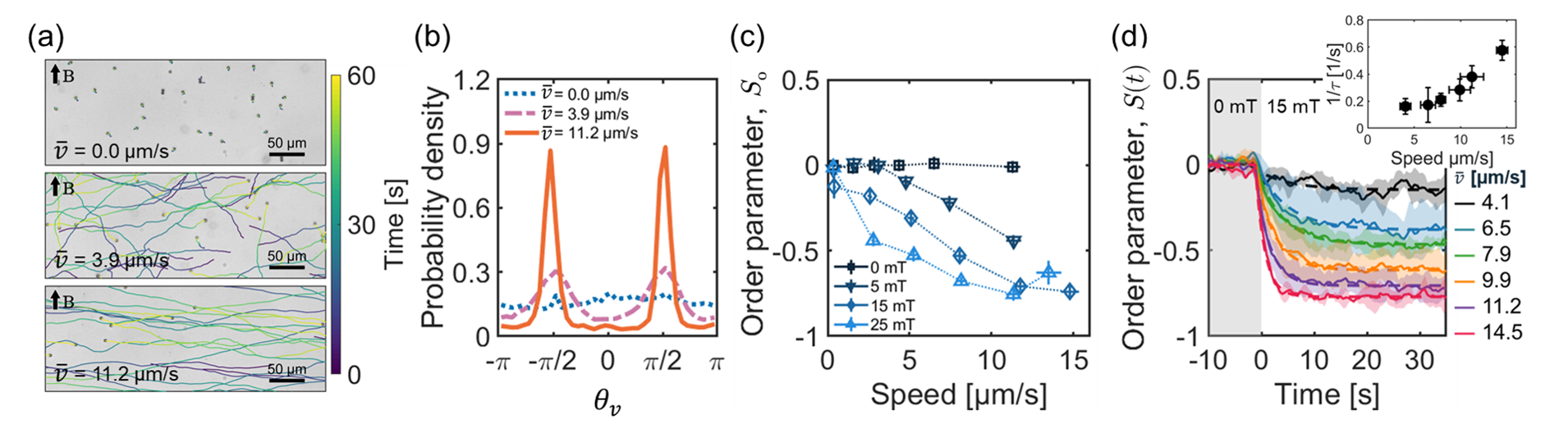}
\caption{(a) Speed dependence of cross-field motion of active Janus colloids under the magnetic field $B=\SI{25}{\milli\tesla}$. (b) Probability distributions of the swimming direction $\theta_v$ for different mean speeds $\bar{v}$. Each distribution was constructed from $N$ tracked particle-velocity data points: $N=29846$ for $\bar{v}=\SI[per-mode=symbol]{0.0}{\micro\meter\per\second}$ ($E=\SI[per-mode=symbol]{0.0}{\volt\per\micro\meter}$), $N=33476$ for $\bar{v}=\SI[per-mode=symbol]{3.9}{\micro\meter\per\second}$ ($E=\SI[per-mode=symbol]{0.033}{\volt\per\micro\meter}$), and $N=17327$ for $\bar{v}=\SI[per-mode=symbol]{11.2}{\micro\meter\per\second}$ ($E=\SI[per-mode=symbol]{0.058}{\volt\per\micro\meter}$). (c) Time-averaged orientational order parameter $S_0$ plotted against the speed $\bar{v}$ for different magnetic flux densities. 
(d) Relaxation dynamics of the orientational order parameter $S(t)$ for different values of the speed $\bar{v}$. Dashed lines are fits to Eq. \eqref{relax}.
Inset: The inverse of response time (1/$\tau$) plotted against the speed $\bar{v}$.}
\label{fig2}
\end{center}
\end{figure*}

To identify the origin of the effective torque responsible for cross-field motion, we examine how the swimming direction depends on propulsion speed, $v$. We tune the propulsion speed by varying the electric field driving ICEP from $E=\SI[per-mode=symbol]{0.008}{\volt\per\micro\meter}$ to \SI[per-mode=symbol]{0.058}{\volt\per\micro\meter} (Fig.~S3). Because individual particles exhibit a distribution of velocities even under a fixed electric-field condition, we characterize each condition by the ensemble-averaged speed, $\bar{v}$. Under a magnetic field of $B=\SI{25}{\milli\tesla}$, particles with a moderate speed ($\bar{v}=\SI[per-mode=symbol]{3.9}{\micro\meter\per\second}$) exhibit preferential orientation at $\theta_v=\pm\pi/2$ with large angular fluctuations (Fig.~\ref{fig2}(a,b)). At a faster speed ($\bar{v}=\SI[per-mode=symbol]{11.2}{\micro\meter\per\second}$), the cross-field alignment becomes sharply peaked (Fig.~\ref{fig2}(a,b)), showing that the stabilization of cross-field alignment depends on self-propulsion strength.

The sharpening of cross-field alignment with increasing propulsion speed suggests that the reorientation mechanism is coupled to self-propulsion. To quantify this tendency, we define the orientational order parameter 
\begin{equation}\label{order1}
    S(t)=\langle \cos2(\theta_{v}^{j}(t) - \theta_B) \rangle_{j},
\end{equation} 
where $\theta_{v}^{j}$ is the heading angle of the velocity of the $j$th particle and $\theta_B=0$ denotes the magnetic-field direction. Here, $S(t)=0$ corresponds to isotropic motion, whereas $S(t)<0$ indicates preferential motion transverse to the field. Because cross-field motion is characterized by the persistent directional bias of the ensemble at the steady state, we quantify it using the time-averaged order parameter, $S_0=\langle S(t)\rangle_t$. Fig.~\ref{fig2}(c) shows that particles exhibit $S_0<0$ under all applied magnetic fields, while $S_0=0$ is observed only at $B=\SI{0}{\milli\tesla}$. Moreover, $|S_0|$ increases with both propulsion speed and magnetic field strength, indicating that stronger \ytm{magnetic} fields more effectively stabilize motion perpendicular to the field.

Having established cross-field motion under an applied magnetic field, we next examine the reorientation dynamics by measuring the response to the field switching. We probe the reorientation dynamics by switching the magnetic field from $B=\SI{0}{\milli\tesla}$ to $B=\SI{15}{\milli\tesla}$ and monitoring the relaxation of $S$ toward its negative steady-state value $S_0$. The relaxation of $S(t)$ is well captured by a single-exponential form~\cite{supplement},
\begin{equation}\label{relax}
S(t)=S_0\left(1-e^{-t/\tau}\right),
\end{equation}
which defines a reorientation timescale $\tau$ (Fig.~\ref{fig2}(d)). The inverse response time $1/\tau$ increases with propulsion speed for $\bar{v}\gtrsim \SI[per-mode=symbol]{5}{\micro\meter\per\second}$ (Fig.~\ref{fig2}(d), inset). This speed dependence is incompatible with a conventional magnetic torque acting directly on a \ytm{non-spherical} particle~\cite{wang2021magnetic,beppu2024magnetically}, and instead points to a hydrodynamic torque generated by self-propulsion in the anisotropic medium.

\begin{figure}[tb!]
    \centering
    \includegraphics[width=\linewidth]{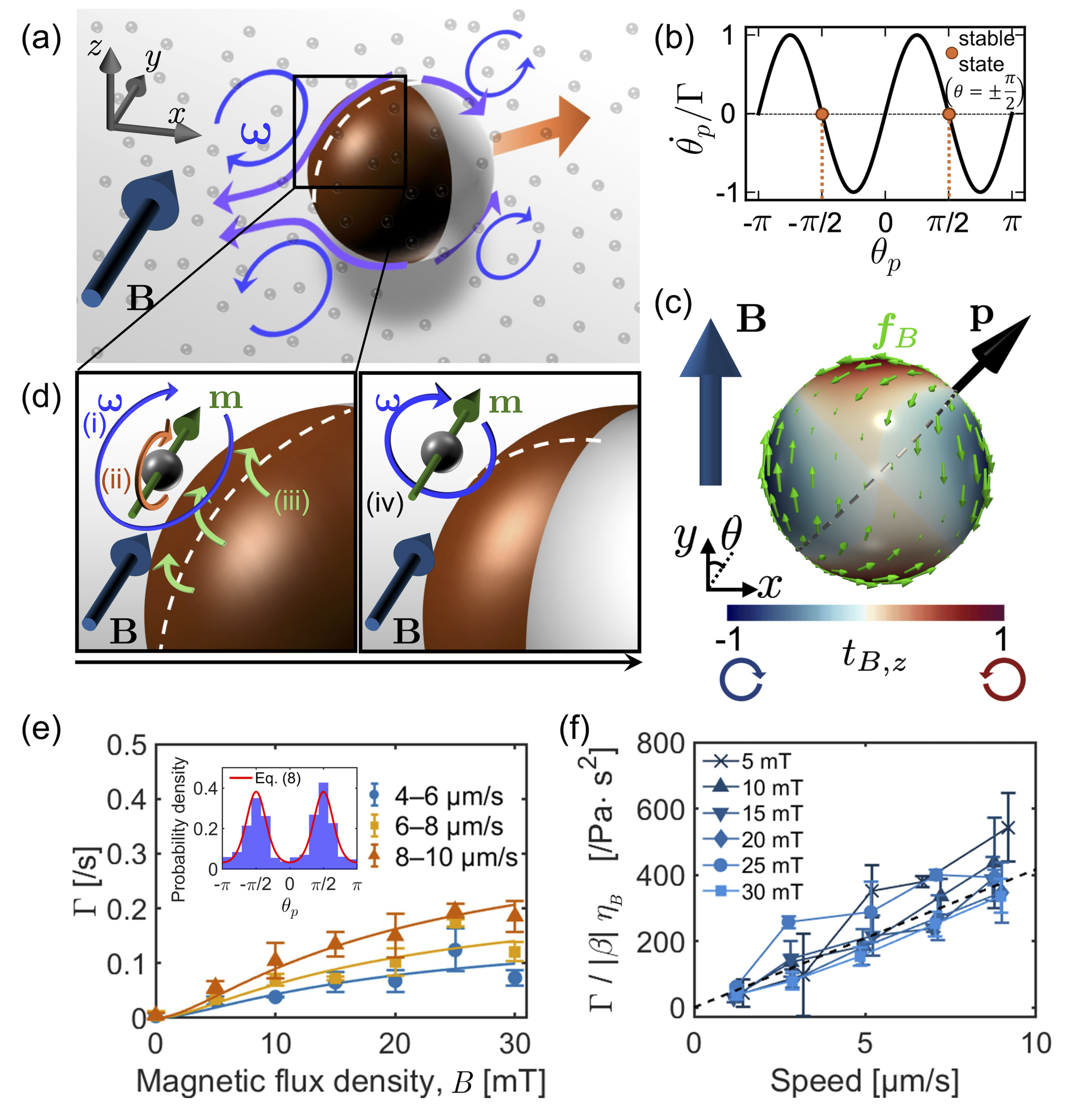}
    \caption{
    (a) Schematic of a pusher-type active colloid in a ferrofluid under a uniform magnetic field $\mathbf{B}$, showing the surface-slip flow and associated vorticity $\bm{\omega}$.
    (b) Angular dependence of the reorientation dynamics $\dot{\theta}_p/\Gamma$, with stable states at $\theta_p=\pm\frac{\pi}{2}$.
    (c) Torque density $t_{B,z}$ on a pusher arising from the stress $\bm{\sigma}_B$. Green arrows on the particle surface indicate the force density $\bm{f}_B$.
    (d) The mechanism of cross-field alignment.
    \added{(i) Pusher-induced vorticity, $\boldsymbol{\omega}$. 
    (ii) Flow-induced rotation of MNP magnetic moment $\mathbf{m}$, counteracted by alignment with $\mathbf{B}$.
    (iii) Magnetoviscous stress: asymmetric surface traction (green arrows).
    (iv) Stable cross-field alignment when the vorticity in the fore-aft plane aligns with the field.} %for $\boldsymbol{\omega}\parallel\mathbf{B}$.}
    (e) The reorientation rate $\Gamma$ plotted against $B$. Particles were grouped into the propulsion-speed ranges. Solid lines are fits to $\Gamma$ in \Eqref{eq:phi_dot}. Inset: Probability density of $\theta_p$ under $B=\SI{30}{\milli\tesla}$ and $v = 4$–\SI{6}{\micro\meter/\second}. The red line is a fit to Eq.~\eqref{FP2}. (f) $\Gamma/|\beta|\eta_B$ against $\bar{v}$ shown for $B=5$ to \SI{30}{\milli\tesla}. The data collapse onto a single master curve (the dashed line).}
    \label{fig3}
\end{figure}

\ytm{To isolate the hydrodynamic mechanism underlying the field-induced reorientation, we consider a minimal model of a spherical swimmer in a ferrofluid with field-induced viscous anisotropy (Fig.~\ref{fig3}(a)). Following Ref.~\cite{Shen2018-bn}, we model the active Janus colloid of radius $a$ as a squirmer~\cite{Blake1971-ws} and retain only the first two polar squirming modes, which correspond to self-propulsion and force-dipole flow.}
The tangential slip velocity profile at the particle surface is prescribed as
\begin{align}\label{eq:slip}
    \bm{u}^\mathcal{S}(\vartheta) = \frac{3}{2}v_0\lb\sin\vartheta + \beta\sin\vartheta\cos\vartheta\rb\bm{e}_\vartheta,
\end{align}
where $v_0$ denotes the propulsion speed, $\vartheta$ is the polar angle measured from the particle orientation vector $\mathbf{p}$, and $\bm{e}_\vartheta$ is the polar unit vector. The squirmer parameter $\beta$ distinguishes pullers ($\beta>0$) from pushers ($\beta<0$)~\cite{Ishikawa2006-hw}. \ytm{The far-field flow of active Janus colloids has been reported to exhibit pusher-like characteristics~\cite{Nishiguchi2015-zw}. We focus on $\beta<0$, throughout the following analysis~\cite{peng2014induced}.} Because the surface slip flow drives active colloids transverse to the applied electric field~\cite{bazant2004induced, Squires2006-zx}, the swimming direction lies in the $xy$ plane. We restrict the analysis to this plane, with $\mathbf{B}$ applied along the $y$ direction. 

Assuming that MNPs remain freely dispersed without chain formation, we describe the ferrofluid stress under a uniform magnetic field by a minimal Shliomis-type constitutive form~\cite{shliomis1974magnetic,Mahle2008-hw}:
\begin{align}\label{eq:stress}
    \bm{\sigma}= -p\bm{1} + \eta_0\dot{\bm{\gamma}} + \underbrace{\eta_B\left[\bm{e}_B(\bm{e}_B\times\bm{\omega}) -(\bm{e}_B\times\bm{\omega})\bm{e}_B\right]}_{\displaystyle \bm{\sigma}_B},
\end{align}
where $\bm{\omega}\equiv \bm{\nabla}\times\bm{u}$ is vorticity, $\bm{u}$ denotes \ytm{the velocity field of the Stokes flow around the squirmer}, and $\bm{e}_B\equiv \mathbf{B}/|\mathbf{B}| (=\bm{e}_y)$ is the magnetic-field direction. The first term is the isotropic pressure enforcing incompressibility, and the second term is the Newtonian viscous stress with shear rate $\dot{\bm{\gamma}}$ and viscosity $\eta_0$. The third term represents the field-induced magnetoviscous stress $\bm{\sigma}_B$,
\kb{which captures the rotational-viscosity response of the ferrofluid arising from the coupling between the local vorticity $\bm{\omega}$ and field-aligned MNPs.}
Its magnitude is set by the magnetoviscous coefficient $\eta_B=\alpha(B)B^2/(4\mu_0)$ where $\mu_0$ is the vacuum permeability and $\alpha(B)$ is the transport coefficient. 

\ytm{To evaluate $\bm{\sigma}_B$ to leading order, we use the Stokes solution obtained for $\eta_B=0$~\cite{Ishikawa2006-hw}, thereby neglecting the feedback of anisotropic viscosity on the flow field~\cite{Lintuvuori2017-hq,supplement}. The torque $\bm{T}$ from $\bm{\sigma}_B$ is then given by,}
\begin{align}\label{eq:torque}
    \bm{T} &= \int_\mathcal{S} {\bm{r}}\times\lb\bm{\sigma}\cdot\hat{\bm{r}}\rb\ dS = \frac{6}{5}\pi\eta_B a^2\beta v_0(\mathbf{p}\cdot\bm{e}_B)(\mathbf{p}\times\bm{e}_B).
\end{align}

The resulting orientational dynamics follow directly from torque balance between Eq.~\eqref{eq:torque} and rotational viscous drag. Let $\theta_p=\cos^{-1}(\mathbf{p}\cdot\bm{e}_B)$ be the polar angle of the squirmer orientation $\mathbf{p}$ measured from $\bm{e}_B$. The corresponding angular velocity about the $z$ axis is $\mathbf{\Omega}=-\dot{\theta}_p\bm{e}_z$. The torque balance relation, $\bm{T}-8\pi\eta_0 a^3\bm{\Omega}=\bm{0}$, yields
\begin{align}\label{eq:phi_dot}
    \dot{\theta}_p = \Gamma\sin(2\theta_p),\quad
    \Gamma = -\frac{3}{40}\frac{\beta v_0}{a}\frac{\eta_B}{\eta_0},
\end{align}  
where $\Gamma$ denotes the reorientation rate ($\Gamma > 0$). \ytm{For a pusher, this equation predicts stable states at $\theta_p=\pm\pi/2$ (Fig.~\ref{fig3}(b)), showing that the torque rotates the particle axis toward an orientation perpendicular to the applied field.} 

\ytm{It is worth noting that the ICEP slip flow of Janus colloids differs from the axisymmetric squirmer slip~\cite{Nishiguchi2015-zw}. 
To test whether this difference affects the reorientation dynamics, we performed direct numerical simulations of spherical swimmers~\cite{Nakayama2005-aa,Molina2013-rw,Yamamoto2021-oe,Kobayashi2023-ad,Kobayashi2024-tn,supplement} separately using the ICEP slip profile and the squirmer slip profile, without relying on the analytical approximations used in the anisotropic stress calculation. 
For both slip profiles in a fluid with uniaxial magnetoviscous anisotropy, the simulations reproduce cross-field motion and agree with the analytical solution of the squirmer model (Fig.~\ref{figS6}). 
This agreement with the analytical squirmer solution supports the squirmer model as an effective hydrodynamic description for the observed reorientation dynamics.} 

To further reveal the origin of the reorientation torque, we illustrate the surface force density $\bm{f}_B$ and the resulting torque density $\bm{t}_B$ arising from the field-induced stress $\bm{\sigma}_B$ (Fig.~\ref{fig3}\ytm{(c)}),
\begin{align}
\bm{f}_B=\left.\bm{\sigma}_B\right|_{r=a}\cdot \hat{\bm{r}},
\qquad
\bm{t}_B=a\,\hat{\bm{r}}\times \bm{f}_B ,
\end{align}
where $\hat{\bm{r}}$ is the radial unit vector. The region with negative $t_{B,z}$ dominates over the positive one, driving pushers toward perpendicular to the applied field (Fig.~\ref{fig3}\ytm{(c)}).

The mechanism of cross-field alignment is summarized as follows (Fig.~\ref{fig3}\ytm{(d)}):
\added{(i) A pusher generates vorticity $\boldsymbol{\omega}$ in its near field.  
(ii) When $\boldsymbol{\omega}$ is not aligned with the magnetic field $\mathbf{B}$, the local rotation tends to misalign the magnetic moment $\mathbf{m}$ of a dispersed MNP with respect to $\mathbf{B}$, while the magnetic torque $\mathbf{m}\times\mathbf{B}$ tends to counteract this flow-induced misalignment. 
(iii) This competition generates a magnetoviscous stress in the surrounding ferrofluid, which produces an asymmetric surface traction on the swimmer (green arrows) and thereby a net reorienting torque.  
(iv) Cross-field alignment arises at the stable orientations where this net torque vanishes.}

\ytm{\Eqref{eq:phi_dot} indicates that the reorientation rate $\Gamma$ increases with the propulsion speed and the field-dependent magnetoviscosity. We next apply a Langevin-type statistical analysis of the orientation dynamics to evaluate $\Gamma$ from the experimental data~\cite{supplement}.} The probability distribution of particle orientation is given by
\begin{equation}\label{FP2}
P(\theta_p) = P_0
\exp\left(-\frac{\Gamma}{D_{\theta}}\cos^2\theta_p\right),
\end{equation}
where $P_0$ is a normalization constant and $D_{\theta}$ is the rotational diffusion coefficient. Using independently measured $D_{\theta}=\SI{0.051}{\per\second}$ (Fig.~S7), fitting Eq.~\eqref{FP2} to the measured $P(\theta_p)$ yields an estimate of $\Gamma$.

The measured $P(\theta_p)$ agrees well with Eq. \eqref{FP2} (Fig.~\ref{fig3}(e) inset). The experimentally obtained $\Gamma$, plotted separately for particles swimming at slow, moderate, and fast speeds, increases with $B$ in all cases (blue, yellow, red in Fig.~\ref{fig3}(e), respectively). 
\added{The data are well described by Eq.~\eqref{eq:phi_dot} using the field-dependent magnetoviscosity $\eta_B$ (see End Matter for details). The gradual saturation of $\Gamma$ at larger $B$ reflects the corresponding field dependence of the magnetoviscous response of the ferrofluid.}

Eq.~\eqref{eq:phi_dot} also predicts that $\frac{\Gamma}{|\beta|\eta_B}$ scales linearly with propulsion speed $v_0$, with a slope of $\frac{3}{40a\eta_0}$. Using the values of $|\beta|\eta_B$ extracted from Fig.~\ref{fig3}(e), we rescale $\Gamma$ and plot it against $\bar{v}$. The data for all magnetic field strengths collapse onto a linear master curve (Fig.~\ref{fig3}(f)), providing further support for the emergent coupling between self-propulsion and anisotropic viscosity.

The aligning mechanism revealed here is structurally related to an emergent torque theoretically predicted for swimmers in liquid-crystalline media~\cite{Lintuvuori2017-hq}: in both cases, swimmer-generated vorticity couples to an anisotropic response of the host medium to generate a reorientation torque. The key distinction is the sign of the anisotropic stress coefficient associated with rotational viscosity, which is positive in the present ferrofluid but negative in the liquid-crystal model~\cite{de1993physics,marenduzzo2007steady}, thereby reversing the stable alignment from perpendicular to parallel. This correspondence suggests that the coupling between swimmer-generated flow and anisotropic rotational viscosity is a generic route to steering self-propulsive motion across microscopically distinct anisotropic media.

\begin{figure}[tb]
\begin{center}
\includegraphics[width=9cm]{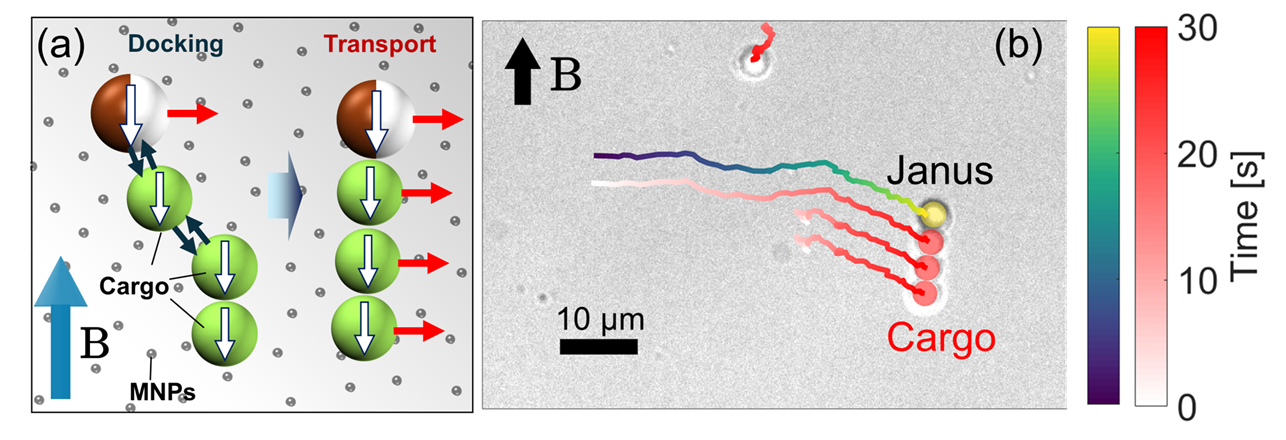}
\caption{(a, b) Cargo particle transport. White arrows and blue arrows in (a) represent the induced magnetic moment and the dipolar attraction, respectively. (b) The active Janus colloid is marked in yellow, and non-magnetic cargo particles are in red. $\Phi=0.24\%$v/v, $B=\SI{30}{\milli\tesla}$, $E=\SI[per-mode=symbol]{0.033}{\volt\per\micro\meter}$. The density of cargo particles is 5.0 particles/\SI{1000}{\micro\meter}$^2$.}
\label{fig4}
\end{center}
\end{figure}

An important consequence of our finding is the ability to load and transport cargo by active matter. To demonstrate this capability, we disperse non-magnetic passive polystyrene (PS) particles with a diameter of \SI{3.0}{\micro\meter} in a ferrofluid containing active Janus colloids. Under an applied magnetic field, both active and passive particles behave as non-magnetic voids in the ferrofluid and acquire induced magnetic moments antiparallel to the field~\cite{Skjeltorp1983Crystallization} (Fig.~\ref{fig4}(a)). The resulting dipolar attraction drives the formation of particle chains aligned along the magnetic field. A Janus particle can dock with PS particles to form composite Janus--cargo assemblies. These assemblies then undergo cross-field motion (Fig.~\ref{fig4}(b), Movie~2), enabling the cargo transport without requiring intrinsic magnetic responsiveness.

In this work, we have demonstrated robust cross-field motion of magnetically inert spherical active colloids in ferrofluids under a uniform magnetic field. Ferrofluids allow externally imposed anisotropic viscosity to be separated from topological defects, elastic effects, and confinement. This separation enables experimental isolation of the hydrodynamic reorientation torque predicted for anisotropic media. More broadly, our results establish anisotropic rheology as a controllable physical variable for converting viscous dissipation into directional symmetry breaking of self-propulsive motion, with dilute dispersions of MNPs providing a minimal experimental realization. The rapid response of ferrofluids further enables reversible control of cross-field alignment within $\tau\sim\SI{1.6}{\second}$ (Fig.~S8, Movie~3), and may provide access to collective motion and pattern formation associated with time-dependent responses and odd viscosity~\cite{markovich2024nonreciprocity,hosaka2024chirotactic}.

\textit{Acknowledgements.---} We would like to thank T. Sakaue and D. Nishiguchi for discussions. This work was supported by Grant-in-Aid for Scientific Research (B) (23K25841), Grant-in-Aid for Transformative Research Areas(B) (26K00015), Grant-in-Aid for Research Activity Start-up (25K23356), Grant-in-Aid for Early-Career Scientists (26K17051), Hosokawa Powder Technology Foundation, the Information Center of Particle Technology, JST FOREST Grant (JPMJFR2239), JSPS Core-to-Core Program “Advanced core-to-core network for the physics of self-organizing active matter" (JPJSCCA20230002). T.K. acknowledges support from the JSPS Overseas Research Fellowship.

\textit{Data availability statement.---} The data that support the findings of this study will be publicly available upon publication. 

\textit{Remark.---} Z.Z. and T.K. equally contributed to this work.

\bibliographystyle{apsrev4-2}
\bibliography{ref}

\begin{center}
{\Large \bf End Matter}
\end{center}

We clarify the quantitative analysis demonstrated in Figs.~\ref{fig3}(e) and \ref{fig3}(f). This analysis provides the field dependence of $\eta_B$ used to test the scaling of the reorientation rate $\Gamma$.

\textit{Magnetic field dependence of $\eta_B$ and $\Gamma$ ---} 
The mass magnetization $M^{*}(B)$ of ferrofluids is well described by the Fr\"ohlich--Kennelly model~\cite{Rigoni2022Ferrofluidic}. Writing the volume magnetization as $M(B)=\rho_{\mathrm f}M^{*}(B)$, we use
\begin{equation}
M^{*}(H)=\frac{M(H)}{\rho_{\mathrm f}}=\frac{\chi_0 H\, M_s}{\rho_{\mathrm f}(\chi_0 H+M_s)},
\qquad
H=\frac{B}{\mu_0},
\label{eq:FK_M_H}
\end{equation}
where $\rho_{\mathrm f}$ is the mass density of the ferrofluid, $M_s$ is the saturation magnetization, $\chi_0$ is the initial magnetic susceptibility, and $\mu_0$ is the vacuum permeability.

We define the field-dependent differential susceptibility by
\begin{equation}
\chi_{\mathrm f}(B)\equiv \frac{\partial M}{\partial H}.
\label{eq:chi_def_end}
\end{equation}
Using Eq.~\eqref{eq:FK_M_H}, this becomes
\begin{equation}
\chi_{\mathrm f}(B)=\frac{\chi_0}{\left(1+\dfrac{B\chi_0}{\mu_0 M_s}\right)^2}.
\label{eq:chi_FK_end}
\end{equation}

For a monodisperse ferrofluid where MNPs remain freely dispersed without chain formation, the transport coefficient $\alpha(B)$ is written as
\begin{equation}
\alpha(B)=\frac{\tau_{\mathrm f}\,\chi_{\mathrm f}(B)}{\left[1+\chi_{\mathrm f}(B)\right]^2},
\label{eq:alpha_chi_sup}
\end{equation}
where $\tau_{\mathrm f}$ denotes the magnetic relaxation time of the free MNPs.
Substituting Eq.~\eqref{eq:chi_FK_end} into Eq.~\eqref{eq:alpha_chi_sup} yields
\begin{equation}
\alpha(B)=\tau_{\mathrm f}\,\chi_0\,
\frac{\left(1+\dfrac{B\chi_0}{\mu_0 M_s}\right)^2}
{\left[\left(1+\dfrac{B\chi_0}{\mu_0 M_s}\right)^2+\chi_0\right]^2}.
\label{eq:alpha_final}
\end{equation}

In addition, using $\eta_B=\alpha(B)B^2/(4\mu_0)$ and Eq.~\eqref{eq:phi_dot} in the main text, the reorientation rate $\Gamma$ is written as
\begin{equation}
\Gamma
=
-\frac{3\,\beta v_0}{160\,a\,\mu_0\,\eta_0}\,
\alpha(B)\,B^2.
\label{eq:Gamma_mid_sup}
\end{equation}
Combining Eq.~\eqref{eq:alpha_final} and Eq.~\eqref{eq:Gamma_mid_sup} gives the detailed expression of $\Gamma$ by
\begin{equation}
\Gamma
=
-\frac{3\,\beta v_0\,\tau_{\mathrm f}\,\chi_0}{160\,a\,\mu_0\,\eta_0}\,
\frac{B^2\left(1+\dfrac{B\chi_0}{\mu_0 M_s}\right)^2}
{\left[\left(1+\dfrac{B\chi_0}{\mu_0 M_s}\right)^2+\chi_0\right]^2}
\label{eq:Gamma_final}
\end{equation}
which is used to describe the magnetic-field dependence of $\Gamma$ for particle groups with a given mean propulsion speed $v_0$. In the analysis of Fig.~3(e), the viscous coefficient $\eta_0=\SI{1.0}{\milli\pascal\second}$, the particle radius $a=\SI{1.8}{\micro\meter}$, $\mu_0$ = $4\pi\times10^{-7}$ H/m, and the magnetometry-determined parameters $\chi_0$ and $M_s$ introduced in the following section are treated as fixed parameters, whereas $\beta\tau_f$ is taken as the fitting parameter.

\textit{Magnetometry-based evaluation of $\chi_0$ and $M_s$ ---} We performed magnetometry of the diluted PBG300 ferrofluid and used the resulting magnetization curve to determine the saturation magnetization $M_s$ and the initial magnetic susceptibility $\chi_0$. In the monodisperse-ferrofluid description in Eq.~\eqref{eq:alpha_final}, $\alpha(B)$ is directly related to the magnetoviscous coefficient $\eta_B$ and is therefore an important quantity for evaluating the master curve in Fig.~\ref{fig3}(f). Because $\alpha(B)$ is expressed as a function of the magnetic susceptibility $\chi_{\mathrm f}$, the parameters that determine its functional form, namely $\chi_0$ and $M_s$, can be estimated from magnetometry-based measurements of the mass magnetization $M^{*}$.

Magnetization measurements were performed at room temperature using a vibrating sample magnetometer (VSM-5, Toei Industry Co., Ltd.) on a diluted PBG300 ferrofluid with a nanoparticle volume fraction of $\Phi=0.12\%$v/v. The magnetic field was swept quasi-statically over the range from $-\SI{0.2}{\tesla}$ to $+\SI{0.2}{\tesla}$.
A volume of \SI{200}{\micro\liter} of the diluted ferrofluid was loaded into a plastic assay tube attached to a quartz rod and sealed to prevent evaporation. 

\begin{figure}[tb]
\begin{center}
\includegraphics[width=7cm]{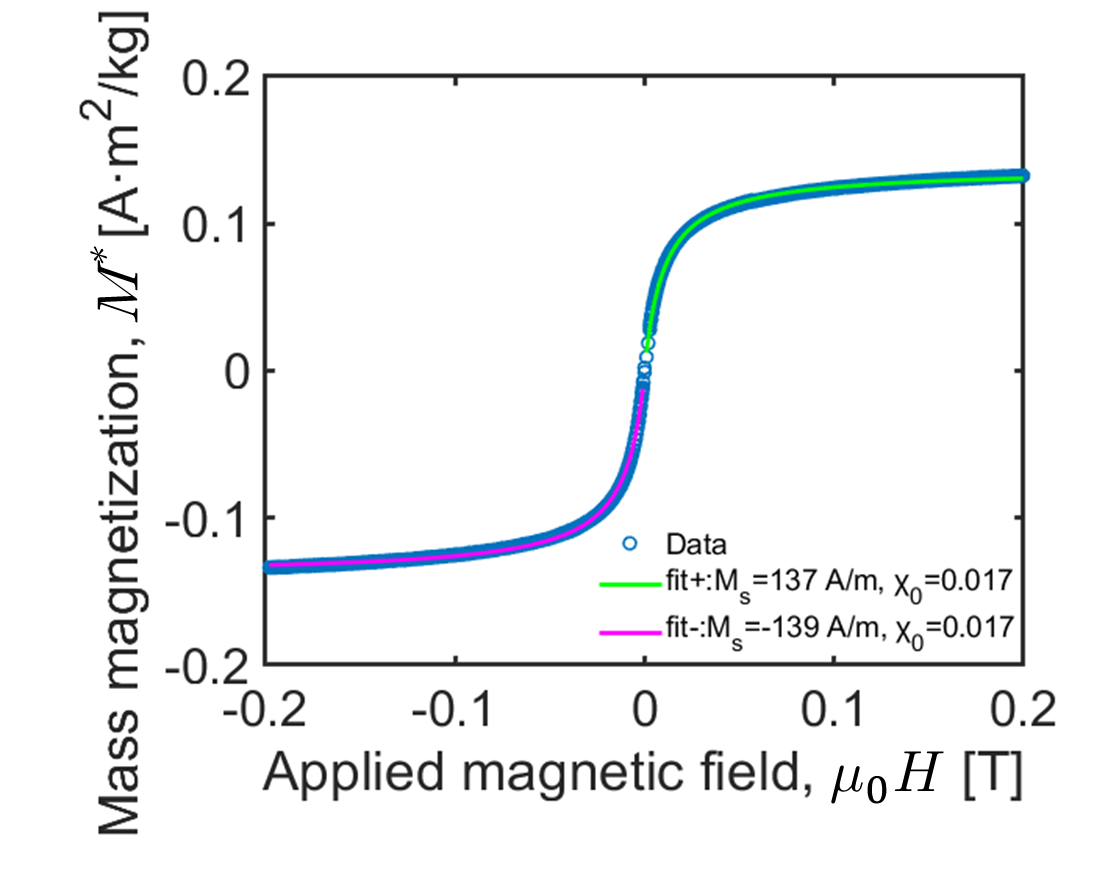}
\caption{Mass magnetization of PBG300 ferrofluids. The positive direction is defined as the magnetic field $\textbf{B}$ parallel to the measurement axis ($\bm{e}_y$), while the negative direction corresponds to the antiparallel field. The fitting curves (green and red) are obtained using Eq.~\eqref{eq:FK_M_H} for $M^*=M/\rho_{\mathrm f}$, with $\rho_{\mathrm f}=\SI[per-mode=symbol]{1.0e3}{\kilogram\per\meter\cubed}$ fixed. The fits yield the saturation magnetization $M_s=\SI[per-mode=symbol]{137}{\ampere\per\meter}$ and $\chi_0 = 0.017$ for the positive-field sweep, and $M_s=\SI[per-mode=symbol]{139}{\ampere\per\meter}$ and $\chi_0 = 0.017$ for the negative-field sweep.}
\label{fig5}
\end{center}
\end{figure}

The measured mass magnetization $M^{*}$ is shown in Fig.~\ref{fig5}. Fitting Eq.~\eqref{eq:FK_M_H} to the positive-field sweep yields $M_s=\SI[per-mode=symbol]{137}{\ampere\per\meter}$  and $\chi_0=0.017$, whereas fitting to the negative-field sweep gives $M_s=\SI[per-mode=symbol]{139}{\ampere\per\meter}$ and $\chi_0=0.017$. The agreement confirms that the Fr\"ohlich--Kennelly form captures the magnetic response of the dilute ferrofluid in the present regime~\cite{Rigoni2022Ferrofluidic}. 
These experimentally determined parameters were further used as material constants when comparing the magnetic-field dependence of $\Gamma$, given by Eq.~\eqref{eq:Gamma_final}, with the data in Fig.~\ref{fig3}(e).

\textit{Evaluation of a linear master curve using Eq.~\eqref{eq:Gamma_final} ---}
We experimentally determined the reorientation rate $\Gamma$ by combining the angular distribution of the particle polarity vector with an independent measurement of the rotational diffusion coefficient. In Fig.~\ref{fig3}(e), $\Gamma$ is plotted against the magnetic field $B$ for three particle groups classified by propulsion-speed range. To compare the field dependence with Eq.~\eqref{eq:Gamma_final}, we assign to each group a representative propulsion speed $v_0$ given by the midpoint of the corresponding speed bin (e.g., $v=4$--\SI[per-mode=symbol]{6}{\micro\meter\per\second} is represented by $v_0=\SI[per-mode=symbol]{5}{\micro\meter\per\second}$), and fit the data for each group separately using $\beta\tau_{\mathrm f}$ as the fitting parameter. The resulting values of $\beta\tau_{\mathrm f}$ vary slightly among the three speed groups, but remain within the range from $-3.58\times10^{-3}$ s to $-4.14\times10^{-3}$ s.

Using these values together with Eq.~\eqref{eq:phi_dot} and Eq.~\eqref{eq:Gamma_final}, we evaluate $|\beta|\eta_B$ separately for each speed group. In Fig.~\ref{fig3}(f), the rescaling is performed using the corresponding bin midpoint as $v_0$ in the theoretical normalization, while the horizontal axis is plotted against the measured mean propulsion speed $\bar{v}$ of each group. Although $\Gamma$ depends on both $B$ and $v_0$, rescaling the independently measured $\Gamma$ by the group-dependent values of $|\beta|\eta_B$ yields a linear master curve that retains only the speed dependence.

\clearpage
\onecolumngrid
\setcounter{equation}{0}
\renewcommand{\theequation}{S\arabic{equation}}
\renewcommand{\thefigure}{S\arabic{figure}}

\begin{center}
{\Large \bf Supplemental Material}
\end{center}
\setcounter{figure}{0}

\section{Materials and Methods}

\subsection{Preparation of Janus colloidal particles}

The Janus colloids were fabricated as follows. A monolayer of silica particles with a diameter of \SI{3.6}{\micro\meter}  (Hyprecica, UEXC) was first formed at an air-water interface and then transferred onto a glass slide. For chromium (Cr)-coated Janus colloids, one surface of the monolayer was first coated with a 20 nm-thick Cr layer using an electron-beam deposition system (EB-20, Vacuum Device). Subsequently, a 5 nm-thick SiO$_2$ layer was deposited on the Cr layer using a sputtering system (nano PVD-S10A, Moorfield Nanotechnology). For gold (Au)-coated Janus colloids, a 3 nm-thick Cr layer was first coated and then followed by a 17 nm-thick Au layer using a thermal deposition system (VPC-260F, ULVAC KIKO). After the deposition, the monolayer on the glass slide was immersed in deionized (DI) water, and the Janus colloids were removed from the glass slide by sonication. The Janus colloids were washed with DI water and isopropyl alcohol. The prepared particles were used as a suspension with a particle density of 0.67$\pm$0.15 particles/\SI{1000}{\micro\meter}$^2$, containing 0.5\%w/v  hydrophilic non-ionic surfactant Pluronic F-127 (P2443, Sigma-Aldrich).

\subsection{Microscopy}
An inverted fluorescence microscope (IX73, Evident) equipped with a CMOS
camera (Neo 5.5, Andor Technology) was used for both phase-contrast and
fluorescence imaging with 20x or 40x objective lenses. Time-lapse recordings of the self-propelled motion
of active Janus colloids were acquired at 1-s intervals with Metamorph software (Molecular Devices). The suspension of the active colloids was introduced into a chamber assembled from two indium tin oxide (ITO)-coated cover slips ($18 \times 18$ mm, resistivity 15–\SI{30}{\ohm}, Alliance Biosystems) separated by a \SI{120}{\micro\meter}-thick double-sided spacer (SecureSeal imaging spacer, Grace Bio-Labs), and an AC voltage was applied through a function generator (maximum output voltage, 10.0 Vpp; AFG-2105, Instek Japan). Before chamber assembly, the ITO-coated cover slips were immersed in a 5.0\%w/v Pluronic F-127 surfactant solution at room temperature for \SI{3}{hours} to prevent non-specific adhesion of the Janus colloids to the ITO electrode surfaces. Polystyrene particles used for demonstrating Janus–cargo assembly transport were fluorescent microspheres with a diameter of \SI{3.0}{\micro\meter} (POL-17147-5, Polysciences, Inc.). For simultaneous imaging of Janus colloids and fluorescent cargo particles, phase-contrast and fluorescence images were acquired
concurrently by combining illumination from two light sources.

\subsection{Image Processing}
Image processing and analysis were performed using MATLAB (MathWorks) and Fiji (open-source software)~\cite{schindelin2012fiji}. Raw microscopy images were preprocessed in MATLAB by binarization and intensity inversion, resulting in images with white particles on a black background. Subsequently, these images were analyzed using the TrackMate plugin~\cite{tinevez2017trackmate} in Fiji.
Particle detection was performed using the Laplacian of Gaussian (LoG) detector, and particle trajectories were reconstructed using the Kalman tracking algorithm to link detections across frames. This procedure yielded the time-dependent positions of individual particles, which were used for subsequent trajectory-based analysis.

\ytm{The polarity vector $\mathbf{p}$ of each Janus particle was determined by the phase contrast image obtained with a 100x objective lens. The Cr-coated hemisphere was first detected from the image contrast. We then identified two characteristic points, (1) the geometric center of the spherical particle, obtained from its circular outline, and (2) the centroid of the image intensity, which reflects the black--white contrast between the Cr-coated hemisphere and the SiO$_2$ hemisphere. The polarity vector $\mathbf{p}$ was defined as the vector from the geometric center to the intensity centroid. The polarity angle $\theta_p$ was then defined as the angle of $\mathbf{p}$ measured from the magnetic-field axis.}

\begin{figure}[tb]
\begin{center}
\includegraphics[width=12cm]{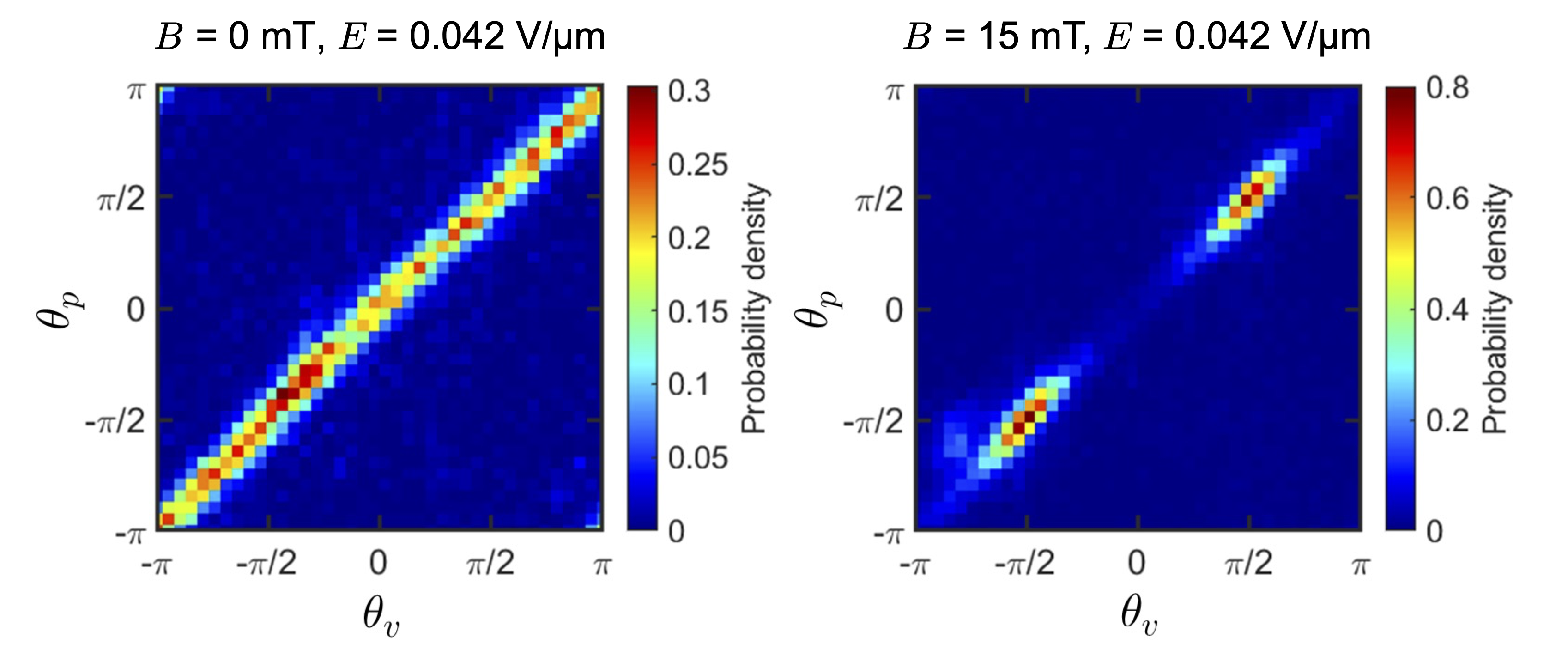}
\caption{Linear correspondence between the direction of the velocity vector $\theta_v$ and the direction of the polarity vector $\theta_p$ of active Janus colloids. The left and right panels correspond to $B=\SI{0}{\milli\tesla}$ and $B=\SI{25}{\milli\tesla}$, respectively.}
\label{figS5}
\end{center}
\end{figure}

\subsection{Ferrofluids and a horizontal uniform magnetic field}
We used a biocompatible ferrofluid (PBG300, Ferrotec) containing sterically stabilized iron-oxide magnetic nanoparticles. The viscosity $\eta_0$ of the diluted ferrofluid containing 0.5\%w/v Pluronic F-127 was measured using a rheometer (MCR 302, Anton Paar) equipped with a cone-and-plate geometry (CP25-1).

A uniform horizontal magnetic field was generated using a pair of custom-made electromagnets (TKS-37032, Toei Co., Ltd.), driven by a DC power supply (PWR401ML, KIKUSUI), with the applied current controlled using sequence creation and control software (SD027-PWR-01, KIKUSUI). The field was uniform over a \SI{5}{\milli\meter}\,$\times$\,\SI{5}{\milli\meter} planar region centered at the sample position. The magnetic field magnitude $B$ was determined from the manufacturer-provided field-current calibration, $B = 68.95\, I - 0.001935$ (\si{\milli\tesla}), where $I$ is the applied current (in \si{\ampere}).

\section{The effect of a magnetic field on particle orientation in the absence of an electric field}

\begin{figure}[tb]
\begin{center}
\includegraphics[width=10cm]{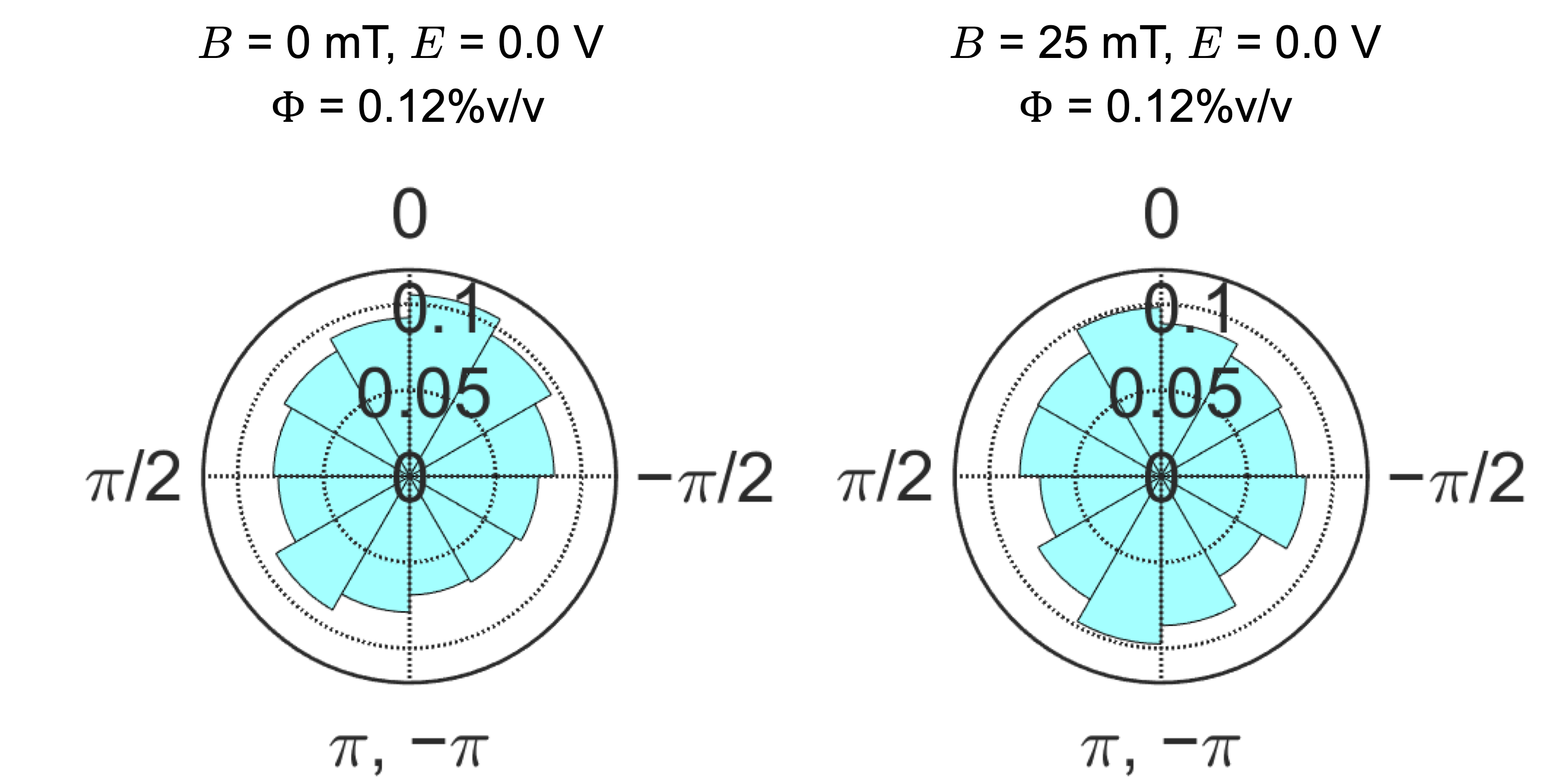}
\caption{Distribution of the orientation angle of Janus colloidal particles in a ferrofluid ($\Phi$= 0.12\%v/v) without AC electric-field driving. (Left) No magnetic field, $B = \SI{0}{\milli\tesla}$ and no electric-field driving, $E=\SI[per-mode=symbol]{0.0}{\volt\per\micro\meter}$. The distribution was constructed from $N=7564$ tracked particle-velocity data points. (Right) With magnetic field $B = \SI{25}{\milli\tesla}$ at the direction of $\theta = 0$ and no electric-field driving $E=\SI[per-mode=symbol]{0.0}{\volt\per\micro\meter}$. $N=8472$.}
\label{figS1}
\end{center}
\end{figure}

To test whether a magnetic field alone can bias the particle orientation, we quantify the orientation-angle distribution of Cr-coated Janus colloids in a ferrofluid ($\Phi = 0.12$\%v/v) without AC electric-field driving. Figure~\ref{figS1} shows that, within the angular range measurable from side-oriented particles, the particles exhibit no orientational preference either in the absence of a magnetic field ($B=\SI{0}{\milli\tesla}$, $E=\SI[per-mode=symbol]{0}{\volt\per\micro\meter}$) or under a uniform magnetic field ($B=\SI{25}{\milli\tesla}$, $E=\SI[per-mode=symbol]{0}{\volt\per\micro\meter}$).
This confirms that the observed cross-field motion does not originate from passive magnetic alignment of the Cr-coated Janus colloids, but requires self-propulsion induced by AC electric-field driving.

\section{Propulsion speed of self-propelled Janus colloidal particles}

\begin{figure}[tb]
\begin{center}
\includegraphics[width=12cm]{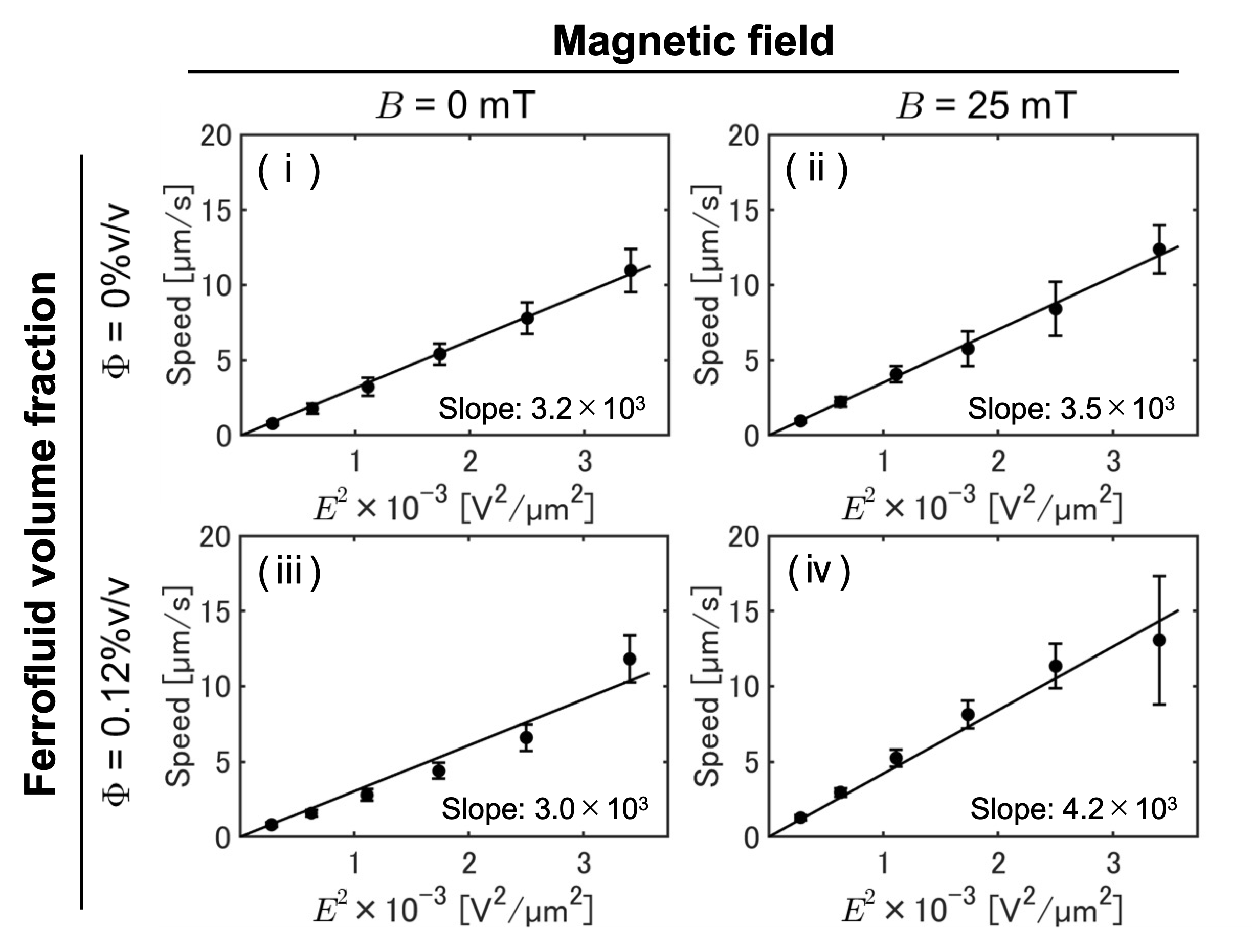}
\caption{The propulsion speed of Cr-coated Janus colloids as a function of $E^2$. The instantaneous velocity $\mathbf{v}=v\mathbf{p}$ was calculated from the trajectories of active colloids moving in either an aqueous solution (($\mathrm{i}$) $\Phi=0\%$v/v, $B=\SI{0}{\milli\tesla}$; ($\mathrm{ii}$) $\Phi=0\%$v/v, $B=\SI{25}{\milli\tesla}$, top-right) or a ferrofluid (($\mathrm{iii}$) $\Phi=0.12\%$v/v, $B=\SI{0}{\milli\tesla}$; ($\mathrm{iv}$) $\Phi=0.12\%$v/v, $B=\SI{25}{\milli\tesla}$). As shown by the fitting curves in each panel, although the slopes slightly differ among the conditions, the mean value of particle speed $\bar{v}$ scales as $E^2$. Error bars represent standard deviation.}
\label{figS2}
\end{center}
\end{figure}

We characterize the self-propelled motion of Cr-coated active Janus colloids by measuring their speed as a function of the applied AC electric field $E$. The particle position is defined as $\bm{r}(t) = (x(t), y(t))$, and the instantaneous velocity as $\bm{v}(t) = \dot{\bm{r}} = d\bm{r}/dt$, estimated from position increments over the time interval of \SI{1.0}{\second}. Using this definition, we compare the ensemble-averaged propulsion speed $\bar{v}$ in an aqueous solution ($\Phi=0\%$v/v) and in a ferrofluid ($\Phi=0.12\%$v/v). For each medium, we further examine both the zero-field condition ($B=\SI{0}{\milli\tesla}$) and the field-applied condition ($B=\SI{25}{\milli\tesla}$) to test whether the propulsion speed scales with $E^2$, as expected for ICEP-driven propulsion. In all cases, the mean speed $\bar{v}$ is found to scale as $E^2$ (Fig.~\ref{figS2}), indicating that the basic ICEP propulsion mechanism is preserved irrespective of the medium and the applied magnetic field.

A comparison of the magnetic-field dependence of the propulsion speed shows that the mean speed is slightly larger for particles undergoing cross-field motion in the ferrofluid under the magnetic field ($\Phi=0.12\%$v/v and $B=\SI{25}{\milli\tesla}$) than under the other conditions (Fig.~\ref{figS2}). We attribute this difference to the reduced orientational fluctuations in the cross-field-aligned state, which promote straighter trajectories and therefore a larger trajectory-based estimate of $\bar{v}$.

\section{Mean-squared displacement}

\begin{figure}[tb]
\begin{center}
\includegraphics[width=18cm]{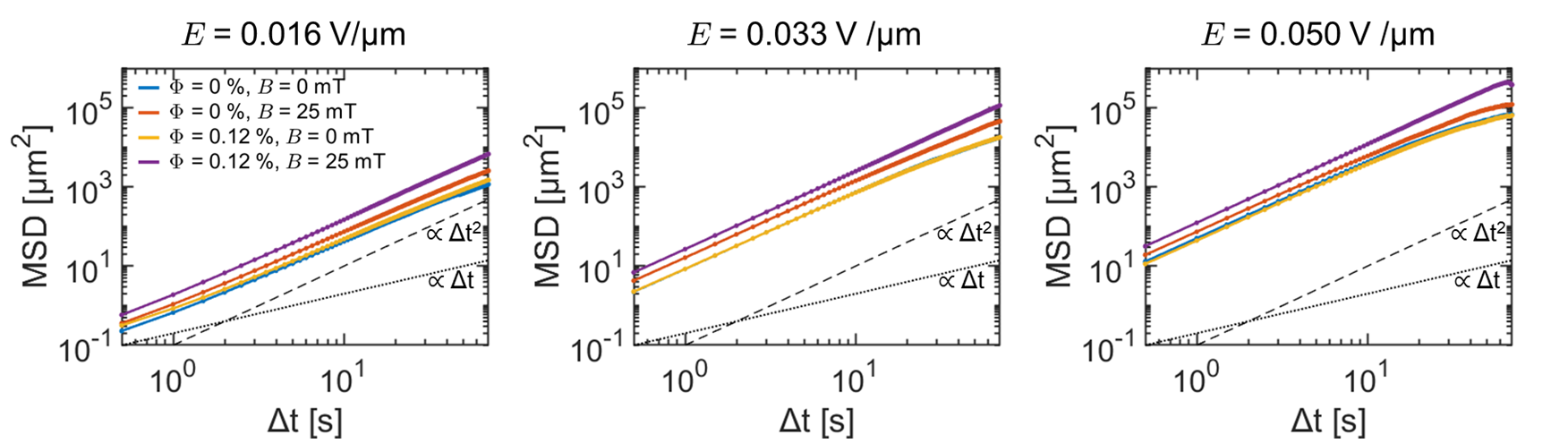}
\caption{Mean-squared displacement (MSD) of active Janus colloids in an aqueous solution and ferrofluid under different magnetic-field conditions. MSD data in the range $t=1$--\SI{20}{\second} were fitted to Eq.~(S1) to determine the power-law exponent $\alpha$. The AC electric field driving the motion of the active colloids was varied as $E=\SI[per-mode=symbol]{0.016}{\volt\per\micro\meter}$ (left), \SI[per-mode=symbol]{0.033}{\volt\per\micro\meter} (center), and \SI[per-mode=symbol]{0.050}{\volt\per\micro\meter} (right). The dashed line indicates ballistic motion ($\mathrm{MSD}\sim t^{2}$), while the dotted line represents normal diffusion ($\mathrm{MSD}\sim t$).
In the aqueous solution without the magnetic field ($\Phi=0\%$v/v, $B=\SI{0}{\milli\tesla}$), the MSD curves exhibit superdiffusive behavior with scaling exponents $\alpha=1.80$, 1.94, and 1.94 for applied AC electric field
of $E=$\SI[per-mode=symbol]{0.016}{\volt\per\micro\meter} (left), \SI[per-mode=symbol]{0.033}{\volt\per\micro\meter} (center), and \SI[per-mode=symbol]{0.050}{\volt\per\micro\meter} (right),
respectively. In the aqueous solution under the magnetic field ($\Phi=0\%$v/v, $B=\SI{25}{\milli\tesla}$), the MSD scalings are $\alpha=1.82$, 1.95, and 1.92 for $E=$\SI[per-mode=symbol]{0.016}{\volt\per\micro\meter} (left), \SI[per-mode=symbol]{0.033}{\volt\per\micro\meter}(center), and \SI[per-mode=symbol]{0.050}{\volt\per\micro\meter} (right),
respectively, remaining unchanged from the zero-field case. 
In contrast, in the ferrofluid without the magnetic field ($\Phi=0.12\%$v/v, $B=\SI{0}{\milli\tesla}$), similar superdiffusive scaling is
observed, with $\alpha=1.72$, 1.93, and 1.94 at $E=$\SI[per-mode=symbol]{0.016}{\volt\per\micro\meter} (left), \SI[per-mode=symbol]{0.033}{\volt\per\micro\meter}(center), and \SI[per-mode=symbol]{0.050}{\volt\per\micro\meter} (right), respectively.
In the ferrofluid under a magnetic field ($\Phi=0.12\%$v/v, $B=\SI{25}{\milli\tesla}$), the MSD approaches ballistic scaling, $\alpha=1.89$, 1.97, and 1.98 for $E=$\SI[per-mode=symbol]{0.016}{\volt\per\micro\meter} (left), \SI[per-mode=symbol]{0.033}{\volt\per\micro\meter} (center), and \SI[per-mode=symbol]{0.050}{\volt\per\micro\meter} (right), respectively. These scaling exponents reflect the suppression of orientational fluctuations due to cross-field alignment.}
\label{figS3}
\end{center}
\end{figure}

We calculate the mean-squared displacement (MSD) of Cr-coated Janus colloids dispersed in an aqueous solution. 

The MSD, $\langle (\Delta \bm{r}(\Delta t))^2 \rangle$, is defined as
\begin{equation}
\langle (\Delta \bm{r}(t))^2 \rangle
= \langle [\bm{r}(t'+t) - \bm{r}(t')]^2 \rangle_{t'}
\propto t^{\alpha},
\end{equation}
where $\langle \cdot \rangle_{t'}$ denotes an ensemble average over the time $t'$.

Both in the presence ($B=\SI{25}{\milli\tesla}$) and absence ($B=\SI{0}{\milli\tesla}$) of a magnetic field, particles driven at $E=\SI[per-mode=symbol]{0.016}{\volt\per\micro\meter}$ exhibit superdiffusive behavior (blue and red curves in Fig.~\ref{figS3}). At higher driving electric fields ($E=\SI[per-mode=symbol]{0.033}{\volt\per\micro\meter}$ and $E=\SI[per-mode=symbol]{0.050}{\volt\per\micro\meter}$), the magnitude of the MSD increases with $E$. In contrast, the superdiffusive exponent $\alpha$ remains nearly unchanged regardless of whether the magnetic field is applied. These results show that, in an aqueous solution, the applied magnetic field does not alter the statistics of the self-propelled motion, consistent with the magnetic inertness of the active Janus colloids.

We next examine the MSD in a ferrofluid at a volume fraction of $\Phi=0.12\%$v/v. In the absence of an applied magnetic field ($B=\SI{0}{\milli\tesla}$), the active colloids still exhibit superdiffusive behavior (yellow curves in Fig.~\ref{figS3}), similar to that observed in the aqueous solution. By contrast, in the ferrofluid under a magnetic field of $B=\SI{25}{\milli\tesla}$, the MSD displays nearly ballistic scaling, with exponents $\alpha = 1.89$–1.97 over all applied AC electric fields (purple curves in Fig.~\ref{figS3}). This ballistic behavior arises because the magnetoviscous torque steers particle motion transverse to the magnetic field, thereby maintaining long-lived directional persistence.

\begin{figure}[tb]
\begin{center}
\includegraphics[width=14cm]{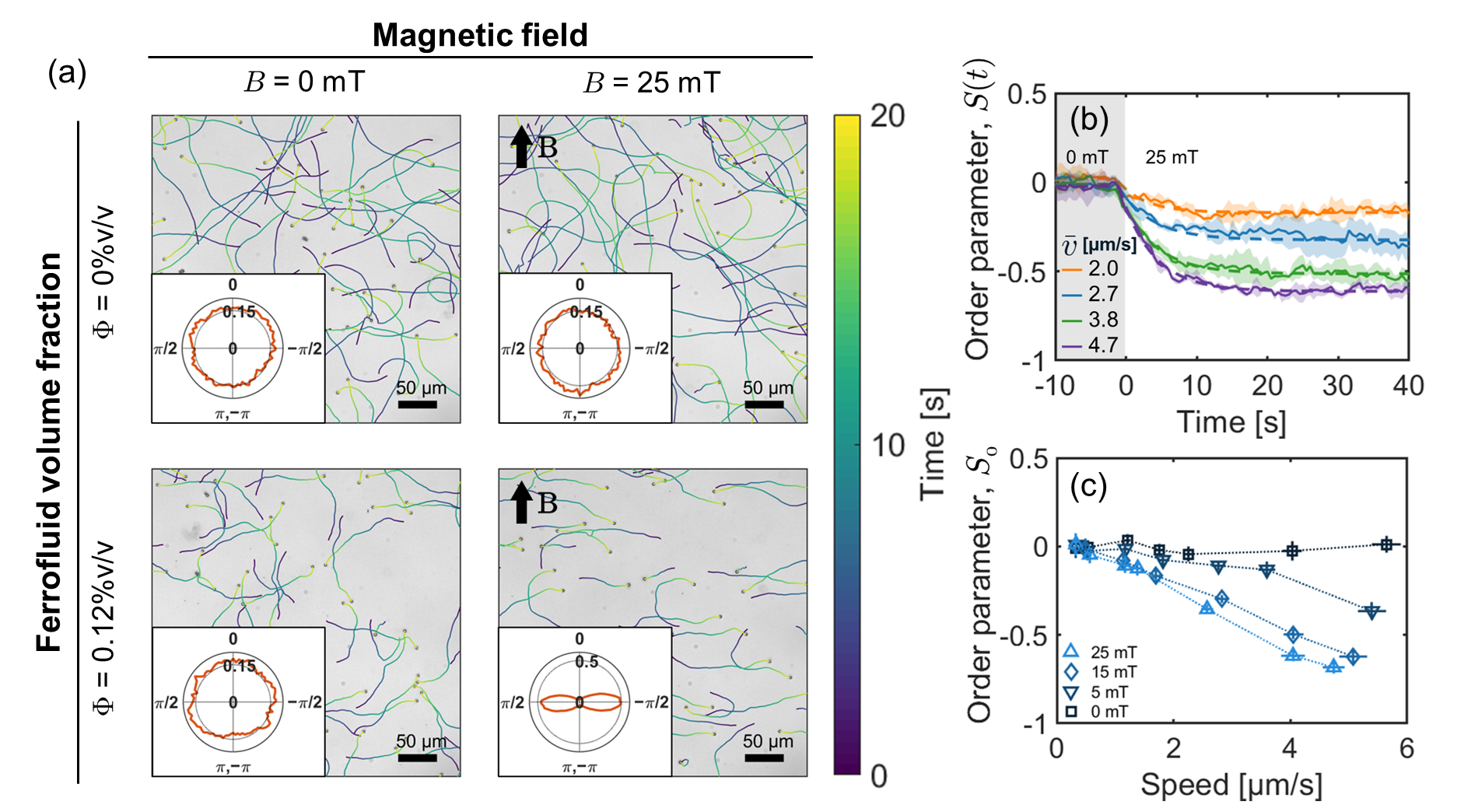}
\caption{(a) Cross-field motion of swimming Au-coated Janus colloids in a ferrofluid. Swimming trajectories are color-coded. 
The uniform magnetic field $B = \SI{25}{\milli\tesla}$ is applied along the $y$ direction ($\theta = 0$). Inset figure: Probability distribution of the angle of particle swimming. Cross-field motion ($\theta_v = \pm \pi/2$) is observed in ferrofluids under the applied magnetic field, as in Cr-coated Janus colloids. Scale bar: \SI{50}{\micro\meter}. (b) The time-averaged orientational order parameter $S_0$ as a function of the propulsion speed $\bar{v}$ for different magnetic flux densities. 
(c) Response to the applied magnetic field. At time $t = 0$, the magnetic field is instantaneously switched from $B = 0$ to $B = \SI{25}{\milli\tesla}$, and the subsequent time evolution of the orientational order parameter $S(t)$ is analyzed.}
\label{figS4}
\end{center}
\end{figure}

\section*{Cross-field motion of Au-coated Janus colloidal particles}

Because Cr-coated Janus colloids do not respond to a magnetic field in the absence of a ferrofluid, the particles themselves are not expected to be directly affected by external magnetic fields with flux densities within $B=$\SI{30}{\milli\tesla}. To examine the dependence on the type of metallic coating, we therefore investigate whether Janus colloids coated with gold (Au) likewise exhibit cross-field motion.

Experiments show that Au-coated Janus colloids at a density
of 0.45$\pm$0.07 particles/\SI{1000}{\micro\meter}$^2$ displayed cross-field motion perpendicular to the applied magnetic field in the ferrofluid of $\Phi =0.12$\%v/v (Fig.~\ref{figS4}(a)-$\mathrm{iv}$). By contrast, they exhibit isotropic motion both in the absence of the ferrofluid, irrespective of the magnetic field (Fig.~\ref{figS4}(a)-$\mathrm{i}$ and $\mathrm{ii}$), and in the ferrofluid when no magnetic field is applied (Fig.~\ref{figS4}(a)-$\mathrm{iii}$). The Au-coated Janus colloids used in this study reach propulsion speeds of \SIrange[per-mode=symbol]{4}{5}{\micro\meter\per\second} under a high applied AC electric field of \SI[per-mode=symbol]{0.090}{\volt\per\micro\meter}. 

We analyze the speed dependence of the cross-field motion using the orientational order parameter 
$S(t)=\langle \cos2(\theta_v^j(t) - \theta_B) \rangle_j$,
where $\theta_v^j(t)$ denotes the heading angle of the velocity of the $j$th particle, $\theta_B=0$ defines the direction of the applied magnetic field, and $\langle \cdot \rangle_j$ denotes an ensemble average over all tracked particles. As in the case of Cr-coated active Janus colloids (Fig.~\ref{fig2}(c) in the main text), the time-averaged orientational order parameter $S_0$ becomes increasingly negative with increasing propulsion speed for Au-coated active Janus colloids (Fig.~\ref{figS4}(b)).

We further examine the response dynamics by applying a step-like magnetic field, switching $B$ from \SI{0}{\milli\tesla} to \SI{15}{\milli\tesla}, and analyzing the resulting reorientation rate. Au-Janus particles respond with the response time being approximately \SI{10}{\second} at $v=4-\SI[per-mode=symbol]{5}{\micro\meter\per\second}$ (Fig.~\ref{figS4}(c)). This behavior is comparable to that observed for Cr-coated Janus colloids swimming at $v=4-\SI[per-mode=symbol]{5}{\micro\meter\per\second}$ (Fig.~2(d) in the main text).

\section*{Derivation of cross-field motion of a squirmer in ferrofluids under a uniform magnetic field}

\tk{
    We consider a spherical squirmer in a ferrofluid under a uniform magnetic field.
    A fully self-consistent analytical solution for this system is not readily available. To obtain a tractable description of the leading-order reorientation mechanism, we therefore introduce the following approximations.
    
    First, we assume that the externally applied magnetic field $\mathbf{B}$ is spatially uniform and defines a fixed anisotropy axis
    \begin{align}
        \bm{e}_B \equiv \mathbf{B}/|\mathbf{B}| \nonumber.
    \end{align}
We neglect field-induced spatial restructuring of MNPs, consistent with the observation that the ferrofluid under a magnetic field does not bias the motion of Janus colloids under the present experimental conditions (Fig.~\ref{figS1}). We instead treat the medium as a homogeneous anisotropic fluid characterized solely by a field-dependent magnetoviscous coefficient $\eta_B$. Under these assumptions, the anisotropic stress of ferrofluids $\bm{\sigma}$ is described by the constitutive form of Mahle \textit{et al.}~\cite{Mahle2008-hw} (Eq.~\eqref{eq:stress} in the main text), in which the field-induced anisotropic stress depends on the local vorticity.}

\ytm{The field-induced stress used in the main text,
\begin{align}\label{sigmaB}
\bm{\sigma}_B
=
\eta_B\left[
\bm{e}_B(\bm{e}_B\times\bm{\omega})
-
(\bm{e}_B\times\bm{\omega})\bm{e}_B
\right],
\end{align}
has an antisymmetric tensor form. This reflects the fact that the magnetoviscous response originates not from ordinary shear deformation, but from a rotational-viscosity mechanism associated with the internal magnetic degrees of freedom of the ferrofluid. In a uniform magnetic field, the magnetic moments of dispersed MNPs tend to align along $\bm{e}_B$. When the surrounding fluid has a local vorticity $\bm{\omega}$, the flow tends to rotate these moments away from the field direction, while magnetic torques act to restore the alignment. In linear response, this competition generates a field-dependent internal torque density proportional to $\bm{e}_B\times\bm{\omega}$. In a coarse-grained continuum description, such an internal torque appears as an antisymmetric contribution to the stress tensor, giving rise to the form above Eq. \eqref{sigmaB}. Thus, the stress $\bm{\sigma}_B$ should be understood as a rotational-viscosity response arising from the finite relaxation of the magnetization under flow.}

Second, following Lintuvuori \textit{et al.}~\cite{Lintuvuori2017-hq}, we evaluate the anisotropic stress using the Stokes flow field of a spherical squirmer in an isotropic fluid. In this approximation, the anisotropy enters only through the constitutive law for the stress, while the velocity field itself is taken to be the standard squirmer solution for Stokes flow field. This corresponds to a leading-order treatment in which the feedback effect of the anisotropic viscosity on the flow field is neglected.

With these assumptions, the velocity field of a squirmer~\cite{Ishikawa2006-hw} is described as
\begin{align}
   \bm{u}(\bm{r}) = \frac{3}{2}v_0\lb\frac{a}{r}\rb^2 \LB\frac{a}{r}\lb\frac{2}{3}\mathbf{p} + \sin\vartheta\bm{e}_{\vartheta}\rb + \frac{\beta}{2}\left\{\lb\frac{a^2}{r^2}-1\rb(3\cos^2\vartheta - 1)\bm{e}_r +\lb\frac{a}{r}\rb^2\sin2\vartheta\bm{e}_{\vartheta} \right\}\RB,
\end{align}
where $\bm{e}_r = \bm{r} / |\bm{r}|$ is the radial unit vector and $\bm{e}_\vartheta$ is the polar unit vector defined with respect to the particle polarity $\mathbf{p}$. The corresponding vorticity field $\bm{\omega}\equiv \bm{\nabla}\times\bm{u}$ is
\begin{align}
    \bm{\omega} = -\frac{9}{2}\beta v_0\frac{a^2}{r^3}(\mathbf{p}\cdot\hat{\bm{r}})\lb\mathbf{p}\times \hat{\bm{r}}\rb.
\end{align}
The field-induced part of the stress, $\bm{\sigma}_B$, gives rise to a traction on the particle surface,
\begin{align}
    \bm{f}_B &= \bm{\sigma}_B|_{r = a}\cdot\hat{\bm{r}}= -\frac{9}{2}\eta_B\frac{\beta v_0}{a}(\mathbf{p}\cdot\hat{\bm{r}})\left[(\mathbf{p}\cdot\hat{\bm{r}})(\bm{e}_B\cdot\hat{\bm{r}})\bm{e}_B - (\bm{e}_B\cdot\mathbf{p})\bm{e}_B - (\bm{e}_B\cdot\hat{\bm{r}})^2\mathbf{p} + (\bm{e}_B\cdot\hat{\bm{r}})(\bm{e}_B\cdot\mathbf{p})\hat{\bm{r}}\right]
\end{align}
with $\eta_B\equiv \frac{\alpha(B)B^2}{4\mu_0}$.

Not surprisingly, the net force vanishes,
\begin{align}
    \bm{F}_B = \int_{\mathcal S}\bm{f}_B\,dS = \bm{0},
\end{align}
as expected in the absence of an external potential. In contrast, the surface traction is spatially asymmetric and therefore generates a finite torque
\begin{align}\label{eq_si:torque}
    \bm{T} &= \int_\mathcal{S} a\hat{\bm{r}}\times\bm{f}_B\ dS = \frac{6}{5}\pi\eta_B a^2\beta v_0(\mathbf{p}\cdot\bm{e}_B)(\mathbf{p}\times\bm{e}_B).
\end{align}
This result shows that the reorientation torque is proportional to both the swimming strength $v_0$ and the anisotropic magnetoviscous coefficient $\eta_B$.

Assuming overdamped rotational dynamics, the reorientation torque is balanced against the rotational viscous drag $8\pi\eta_0 a^3\bm{\Omega}$. We then obtain the angular velocity induced by the anisotropic stress
\begin{align}\label{eq:omega}
    \bm{\Omega} = \frac{3}{20}\frac{\beta v_0}{a}\frac{\eta_B}{\eta_0}(\mathbf{p}\cdot\bm{e}_B)(\mathbf{p}\times\bm{e}_B),
\end{align}

We now restrict the motion to the $xy$ plane, consistent with the experimental geometry. We define the polarity angle of the particle orientation $\theta_p$ from the magnetic-field axis $\bm{e}_B$, defined as $\theta_p \equiv \cos^{-1}(\mathbf{p}\cdot\bm{e}_B)$. 
The equations of motion then read
\begin{align}\label{orientation1}
    \dot{x} = v_0\sin\theta_p,\quad\dot{y} = v_0\cos\theta_p,\quad \dot{\theta}_p = \Gamma\sin(2\theta_p),
\end{align}
where
\begin{align}\label{eq:Gamma}
    \Gamma = -\frac{3}{40}\frac{\beta v_0}{a}\frac{\eta_B}{\eta_0}.
\end{align}
For a pusher ($\beta<0$), $\Gamma>0$ and the stable orientations are $\theta_p=\pm\pi/2$, corresponding to cross-field motion.

\begin{figure}[tb]
\centering
\includegraphics[width=0.8\linewidth]{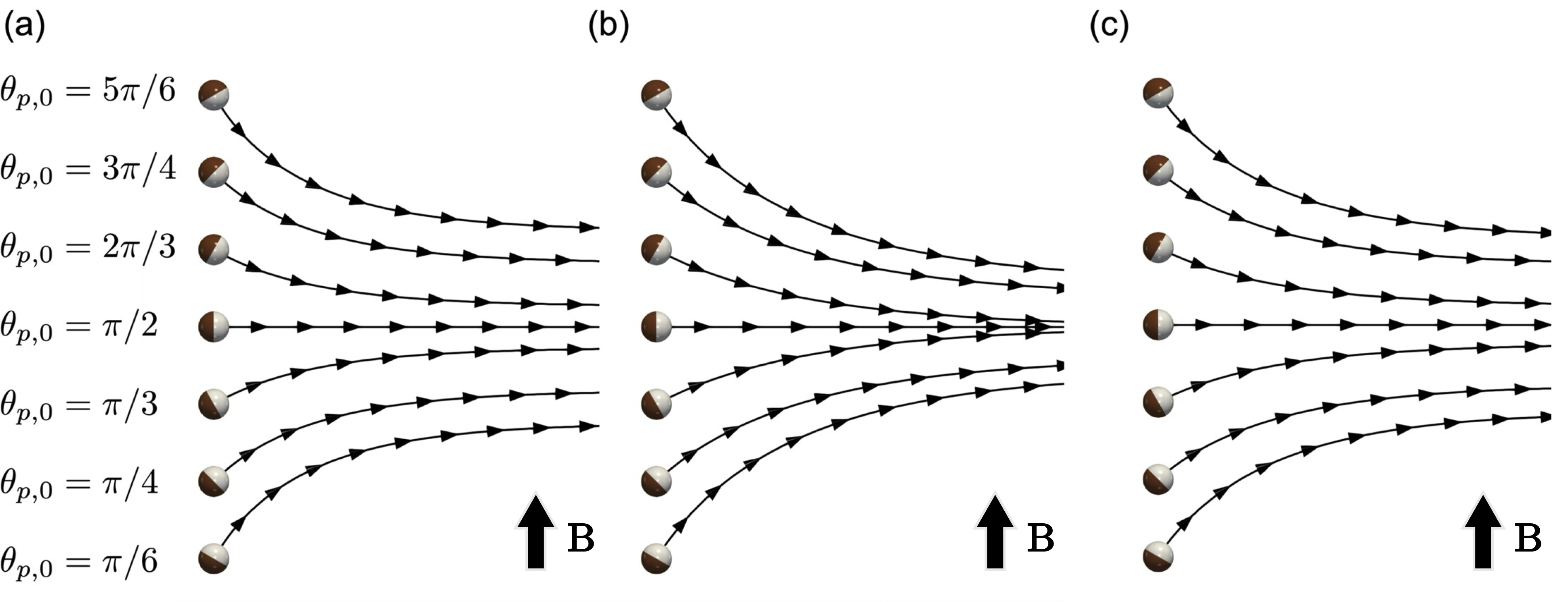}
\caption{
    Cross-field motion of active particles in ferrofluids for $\eta_B/\eta_0 = 0.5$ obtained from analytical solution and direct numerical simulations. The magnetic field is applied along the $y$ axis (upward).
    (a) Analytical solution for the squirmer model with $\beta = -1$.
    (b) Direct numerical simulation of the squirmer model with $\beta = -1$.
    (c) Direct numerical simulation of ICEP-driven active Janus colloids. Trajectories of active particles with different initial orientations $\theta_{p,0}$. All trajectories converge to a direction perpendicular to the magnetic field $\mathbf{B}$.
}
\label{figS6}
\end{figure}

Eqns. \eqref{orientation1} can be solved analytically as
\begin{align}
\begin{split}
    x(t) &= \frac{v_0}{4\Gamma}\ln \LB\left|\frac{1 + \sin\theta_p}{1-\sin\theta_p}\right|\left|\frac{1 - \sin\theta_{p,0}}{1+\sin\theta_{p,0}}\right|\RB + x_0\\
    y(t)&= \frac{v_0}{4\Gamma}\ln \LB\left|\frac{1 - \cos\theta_p}{1+\cos\theta_p}\right|\left|\frac{1 + \cos\theta_{p,0}}{1-\cos\theta_{p,0}}\right|\RB + y_0
    \\
    \tan\theta_p(t) &= \tan\theta_{p,0}e^{2\Gamma t}
\end{split}
\end{align}
where $\theta_{p,0}$ is the initial orientation. The trajectories of squirmer particles are shown in Fig.~\ref{figS6}(a).

We emphasize that this derivation is intended to isolate the leading-order hydrodynamic mechanism of cross-field motion experimentally observed. In particular, it neglects the feedback of the anisotropic viscosity on the velocity field itself. This higher-order effect may modify quantitative coefficients, but the present minimal treatment captures the directional symmetry breaking of spherical Janus colloids and the dependence of the emergent reorientation torque on both self-propulsive motion (propulsion speed $v_0$) and anisotropic rheology in bulk fluid (magnetoviscous coefficient $\eta_B$) in Eq. \eqref{eq:Gamma}.

\section*{Direct numerical simulations of an ICEP Janus particle in ferrofluids}
We explicitly solve for an ICEP Janus colloidal particle in ferrofluids through direct numerical simulations. To do this, we employ the smoothed profile method~\cite{Nakayama2005-aa, Yamamoto2021-oe}. In this method, the sharp boundary between the particle and the surrounding fluid is replaced by a continuous diffuse interface of thickness $\xi$ through a smoothed profile function $\phi \in [0, 1]$, which takes the value $\phi = 1$ inside the particle and $\phi = 0$ in the fluid domain. The mathematical definition of $\phi$ is given in Ref.~\cite{Kobayashi2023-ad}.
The total velocity $\bm{u} = \bm{u}_f + \bm{u}_p$, consisting of the host fluid $\bm{u}_f$ and particle contributions $\bm{u}_p$, evolves according to a modified Navier--Stokes equation,
\begin{align}
    \bm{\nabla}\cdot\bm{u} = 0,\quad \rho\lb\partial_t + \bm{u}\cdot\bm{\nabla}\rb\bm{u} = \bm{\nabla}\cdot\bm{\sigma} + \rho\phi(\bm{f}_p + \bm{f}_{s})
\end{align}
Here, $\rho$ is the fluid density and $\bm{\sigma}$ is the fluid stress tensor, given by \Eqref{eq:stress} in the main text.
The body force $\bm{f}_p$ enforces the rigidity of the particle, whereas $\bm{f}_s$ represents the force distribution associated with the prescribed slip-velocity $\bm{u}^\mathcal{S}$~\cite{Molina2013-rw}. This formulation has been successfully applied to non-Newtonian solvents~\cite{Kobayashi2023-ad, Kobayashi2024-tn}.
The translational and rotational motions of the spherical particle are governed by the Newton--Euler equations,
\begin{align}
    \dot{\bm{R}} &= \bm{V},& \dot{\mathbf{p}}&=\bm{\Omega}\times\mathbf{p}\\
    M\dot{\bm{V}} &= \bm{F}^{\rm H} + \bm{F}^{\rm slip},& \bm{I}\cdot\dot{\bm{\Omega}}_i &= \bm{T}^{\rm H} + \bm{T}^{\rm slip}
\end{align}
 where $\bm{R}$ and $\bm{V}$ denote the position and velocity of the particle, respectively, $\mathbf{p}$ is the particle orientation, and $\bm{\Omega}$ is the angular velocity. $\bm{F}^{\rm H}$ and $\bm{T}^{\rm H}$ are the hydrodynamic force and torque, whereas $\bm{F}^{\rm slip}$ and $\bm{T}^{\rm slip}$ denote the force and torque arising from the imposed slip motion.

We performed simulations for both a single squirmer, whose slip velocity is given by Eq.~\eqref{eq:slip} in the main text, and an ICEP Janus colloid. For the Janus particle, the ICEP slip velocity~\cite{Squires2006-zx, Nishiguchi2015-zw} is given by,
\begin{align}\label{eq:ICEP_slip}
    \bm u^\mathcal{S}=
    \begin{dcases}
        0& (0 \le \vartheta \le \pi/2)\\
        -\frac{9}{8}U_0\lb\sin2\vartheta\cos^2\varphi\bm{e}_\vartheta - \sin\vartheta\sin2\varphi\bm{e}_\varphi\rb
        & (\pi/2 < \vartheta \le \pi).
    \end{dcases}
\end{align}
The particle radius was set to $a=5\Delta$, and the interface thickness was $\xi=2\Delta$, where $\Delta$ is the grid spacing. The particle was placed in a cubic simulation box of side length $L=128\Delta$, with periodic boundary conditions imposed in all directions. Previous studies using the smoothed profile method have shown that this particle resolution is sufficient to accurately capture non-Newtonian hydrodynamic effects around squirmers~\cite{Kobayashi2024-tn}.

Both the squirmer slip (\Eqref{eq:slip} in the main text) and the ICEP slip (\Eqref{eq:ICEP_slip}) yield qualitatively identical reorientation in the simulations (\Figref{figS6}(b) and \Figref{figS6}(c), respectively). Trajectories initialized at various orientations converge to motion transverse to $\mathbf{B}$. This confirms that the squirmer model captures the essential physics of the Janus ICEP system.

\section*{Derivation of the steady-state orientational distribution}
The deterministic reorientation dynamics of a squirmer in a ferrofluid
is described by Eqs.~\eqref{orientation1} and \eqref{eq:Gamma},
\begin{equation}
\dot{\theta}_p=\Gamma \sin(2\theta_p),
\qquad
\Gamma=-\frac{3}{40}\frac{\beta v_0}{a}\frac{\eta_B}{\eta_0}.
\label{eq:theta_det}
\end{equation}

In experiments, the particle orientation fluctuates due to rotational
noise. We therefore rewrite Eq.~(\ref{eq:theta_det}) with an additive
Gaussian white-noise term $\xi(t)$ and write a Langevin-type equation,
\begin{equation}
\dot{\theta}_p=\Gamma \sin(2\theta_p)+\xi(t).
\label{eq:theta_langevin}
\end{equation}
The noise satisfies $\langle \xi(t)\rangle=0$ and
$\langle \xi(t)\xi(t')\rangle = 2D_{\theta}\,\delta(t-t')$, where
$D_{\theta}$ is the rotational diffusion coefficient and $\delta(\cdot)$
denotes the Dirac delta function. Eq.
(\ref{eq:theta_langevin}) thus describes overdamped angular dynamics in
an effective drift field $\Gamma\sin(2\theta_p)$.

The corresponding Fokker–Planck equation for the probability density
$P(\theta_p,t)$ follows from the standard expansion as
\begin{equation}
\frac{\partial P(\theta_p,t)}{\partial t}
= -\Gamma\frac{\partial}{\partial \theta_p}
\!\left[(\sin2\theta_p)P(\theta_p,t)\right]
+ D_{\theta}\frac{\partial^2 P(\theta_p,t)}{\partial \theta_p^2}.
\label{eq:FP_specific}
\end{equation}

To obtain the steady-state distribution $P(\theta_p)$, we set
$\partial_t P(\theta_p,t)=0$ and impose the condition of vanishing
probability flux for a periodic angular variable. This yields
\begin{equation}
\frac{\partial P}{\partial \theta_p}
=\frac{\Gamma}{D_{\theta}}\sin(2\theta_p)\,P(\theta_p).
\label{eq:ODE}
\end{equation}
Integrating Eq.~(\ref{eq:ODE}) gives
\begin{equation}
\ln P(\theta_p)
= -\frac{\Gamma}{2D_{\theta}}\cos(2\theta_p)+\mathrm{const}.
\end{equation}
Using $\cos(2\theta)=2\cos^2\theta-1$, the constant term
can be absorbed into the normalization. We then obtain
\begin{equation}\label{FPfinal}
P(\theta_p) = P_0
\exp\!\left(-\frac{\Gamma}{D_{\theta}}\cos^2\theta_p\right),
\end{equation}
which corresponds to Eq.~(\ref{FP2}) in the main text. The normalization constant $P_0$ is determined by
$\int_0^{2\pi} P(\theta_p)\,d\theta_p = 1$.

\begin{figure}[b]
\begin{center}
\includegraphics[width=12cm]{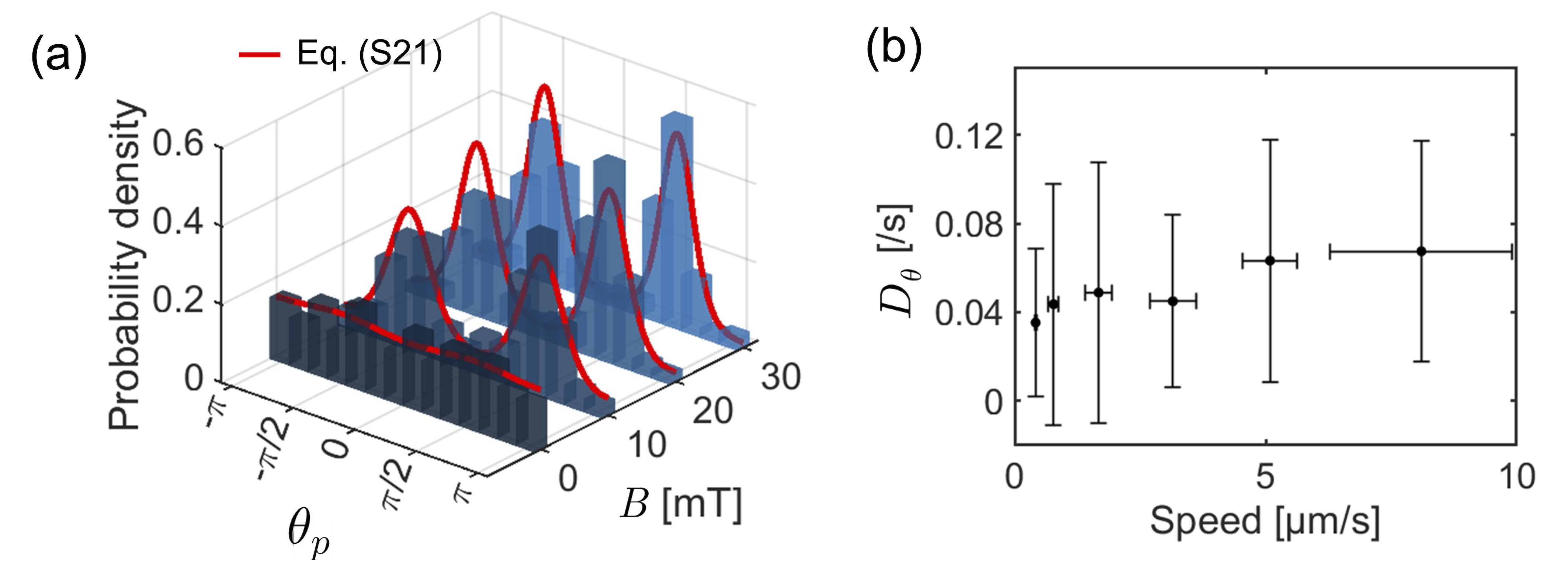}
\caption{(a) Probability density of $\theta_p$ under various magnetic fields. The range of propulsion speed is $v = 4$–\SI{6}{\micro\meter/\second}. Red line is fit to Eq. \eqref{FPfinal}. (b)Rotational diffusion constant $D_{\theta}$ as a function of $\bar{v}$. Error bars represent the standard deviation.}
\label{figS7}
\end{center}
\end{figure}

By measuring the orientational angle of active colloids and fitting the resulting distribution with Eq.~\eqref{FPfinal}, the ratio $\Gamma/D_{\theta}$ can be extracted (Fig.~\ref{figS7}(a)). In addition, the rotational diffusion coefficient $D_{\theta}$ is determined independently from the trajectories of active Janus colloids at field-off and torque-free conditions (Fig.~\ref{figS7}(b)). Specifically, the autocorrelation function of the swimming direction $\theta_v(t)$ was calculated as
 $\left\langle \cos\left[\theta_v(t'+t)-\theta_v(t')\right]
\right\rangle_{t'}=
\exp(-D_{\theta}t)$ where $t$ is the time lag. We extracted $D_{\theta}$ from the autocorrelation function calculated for each particle trajectory and then computed the mean value for each speed range. We used their average over all speed ranges, $D_{\theta}=\SI{0.051}{\per\second}$, for the quantitative estimation of $\Gamma$.

\section*{Linearized relaxation dynamics of the orientational order parameter $S(t)$}

We show that, in the presence of rotational noise, the order parameter $S(t)$ obeys a single-exponential relaxation equation in the vicinity of the stable orientations. For pushers ($\beta<0$), one has $\Gamma>0$, and the stable orientations are $\theta_p=\pm \pi/2$.

To describe the long-time relaxation toward one of these stable states, we expand $\theta_p$ as $\theta_p=\pi/2+\delta\theta_p$, where $\delta\theta_p$ is an infinitesimal deviation under a large reorientation rate ($D_{\theta}\ll \Gamma$). We then obtain $\sin(2\theta_p)\simeq -2\delta\theta_p$, and Eq.~\eqref{eq:theta_langevin} reduces to the Ornstein--Uhlenbeck form
\begin{equation}
\frac{d\delta\theta_p}{dt}=-2\Gamma\delta\theta_p+\xi(t).
\label{eq:delta1_eq}
\end{equation}

The corresponding equation for the second moment is
\begin{equation}
\frac{d}{dt}\langle (\delta\theta_p)^2\rangle = -4\Gamma \langle (\delta\theta_p)^2\rangle + 2D_\theta.
\label{eq:delta2_eq}
\end{equation}

Next, we relate this fluctuation magnitude to the orientational order parameter
\begin{equation}
S(t)=\langle \cos 2(\theta_{p}(t)-\theta_B)\rangle.
\label{eq:S_linear}
\end{equation}
Here we use the fact that, for the active colloids considered here, the swimming direction $\theta_v$ coincides with the angle of the particle polarity vector $\theta_p$ (Fig. \ref{figS5}). The direction of the uniform magnetic field is defined as $\theta_B=0$. The notation $\langle \cdot \rangle$ denotes the ensemble average over the particle population. 

Using the expansion $\cos 2\theta_p\simeq -(1-2(\delta\theta_p)^{2})$, Eq.~\eqref{eq:S_linear} becomes
\begin{equation}
S(t)\simeq -1 + 2\langle (\delta\theta_p)^2\rangle.
\label{eq:S_delta_relation}
\end{equation}
Substituting Eq.~\eqref{eq:delta2_eq} into the time derivative of Eq.~\eqref{eq:S_delta_relation}, and eliminating $\langle (\delta\theta_p)^2\rangle$ using Eq.~\eqref{eq:S_delta_relation}, we arrive at the relaxation equation
\begin{equation}
\frac{dS}{dt} = -4\Gamma (S-S_0),
\label{eq:S_relax_linear1}
\end{equation}
with the steady-state value $S_0=-1+D_\theta/\Gamma$ under the linearized approximation.

We find that Eq.~\eqref{eq:S_relax_linear1} has the solution
\begin{align}
S(t)=S_0+\left[S(0)-S_0\right]e^{-4\Gamma t}.
\label{eq:S_solution_linear}
\end{align}
For an initially isotropic state, $S(0)\simeq 0$, this becomes
\begin{equation}
S(t)=S_0\left(1-e^{-\frac{t}{\tau}}\right),
\label{eq:S_exp_linear}
\end{equation}
with the characteristic response time
\begin{equation}
\tau=\frac{1}{4\Gamma}.
\label{eq:tau_linear}
\end{equation}
Thus, in the vicinity of the stable orientations, the noisy orientational dynamics predicts a single-exponential relaxation of the orientational order parameter. Eqs.~\eqref{eq:S_exp_linear} and~\eqref{eq:tau_linear} therefore explain the exponential relaxation in Eq.~(2) of the main text. 

In addition, using the expression of $\Gamma$ in Eq.~\eqref{eq:Gamma},
the steady-state value $S_0=-1 + D_{\theta}/\Gamma$ for pushers ($\beta<0$) can be written as
\begin{equation}
S_0 = -1+\frac{40a\eta_0D_\theta}{3|\beta|v_0\eta_B}.
\label{eq:S0_linear_pusher}
\end{equation}
This expression shows that, within the linearized approximation, $S_0$ approaches $-1$ as the propulsion speed $v_0$ or the magnetoviscous coefficient $\eta_B$ increases, whereas larger rotational diffusion $D_\theta$ weakens the alignment. Thus, faster propulsion and stronger anisotropic viscosity can enhance the steady cross-field alignment. Figs.~\ref{fig2}(c, d) in the main text also show that the deviation of $S_0$ from $-1$ decreases as the propulsion speed $\bar{v}$ increases. This trend is consistent with Eq.~\eqref{eq:S0_linear_pusher}.

\section*{Rapid and reversible control of cross-field motion of Janus colloidal particles}

\begin{figure}[htb]
\begin{center}
\includegraphics[width=17cm]{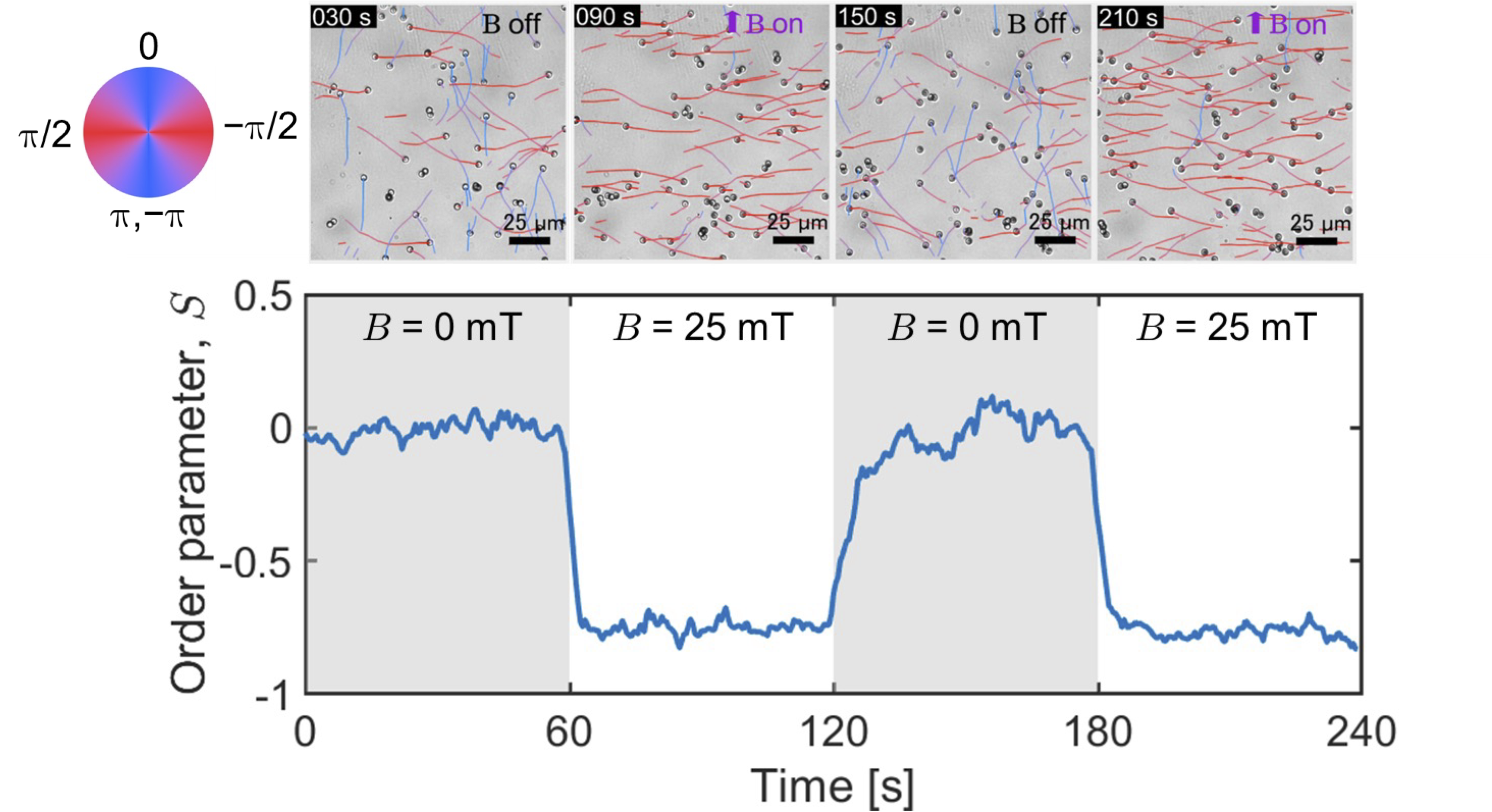}
\caption{(a) Reversible switching of cross-field motion induced by turning the magnetic field ($B=\SI{25}{\milli\tesla}$) on and off. The driving AC field is $E=\SI[per-mode=symbol]{0.058}{\volt\per\micro\meter}$, and the propulsion speed $\bar{v}=\SI[per-mode=symbol]{10.5}{\micro\meter\per\second}$. Colored lines in the microscopy images indicate the instantaneous orientation of the particles. The magnetic field is applied along the $y$ direction (vertical axis).
(b) Time evolution of the orientational order parameter $S(t)$. The magnetic field is alternately switched on and off at \SI{60}{\second} intervals. The orientational order parameter alternates between values close to $S=0$ and $S=-0.8$, and the response time $\tau$ is within \SIrange{2}{3}{\second}, demonstrating reversible and repeatable switching of cross-field alignment. }
\label{figS9}
\end{center}
\end{figure}

To demonstrate reversible control of cross-field motion, we alternately switch the magnetic field on and off at \SI{60}{\second} intervals and quantify the response using the orientational order parameter $S(t)$. Representative microscopy images show isotropic motion in the field-off state and pronounced cross-field motion with the propulsion speed $\bar{v}=\SI[per-mode=symbol]{10.5}{\micro\meter\per\second}$ in the field-applied condition (Fig.~\ref{figS9}(a)). The corresponding time trace of $S(t)$ confirms that cross-field alignment is established within \SIrange{2}{3}{\second} after field activation and relaxes back to the isotropic state within a similar timescale after the field is turned off (Fig.~\ref{figS9}(b)). This fast on–off response, comparable to the rotational diffusion timescale of the Janus colloids, demonstrates rapid and reversible magnetic control of the swimming direction.

\section{Supplemental Movies}

\begin{itemize}
\item Movie 1: Cross-field motion of active Janus colloids in a ferrofluid under an applied magnetic field, corresponding to condition ($\mathrm{iv}$) $\Phi=0.12\%$v/v and $B=\SI{25}{\milli\tesla}$. The magnetic field is applied along the $y$ direction (vertical axis). Under the other conditions, ($\mathrm{i}$) $\Phi=0.0\%$v/v and $B=\SI{0}{\milli\tesla}$, ($\mathrm{ii}$) $\Phi=0.0\%$v/v and $B=\SI{25}{\milli\tesla}$, and ($\mathrm{iii}$) $\Phi=0.12\%$v/v and $B=\SI{0}{\milli\tesla}$, the active Janus colloids exhibit self-propelled motion with no preferred orientation. The ICEP driving condition is $E=\SI[per-mode=symbol]{0.058}{\volt\per\micro\meter}$ at 2~kHz. \\

\item Movie 2: Transport of polystyrene cargo particles by an active Janus colloid showing cross-field motion ($\Phi=0.24$\%v/v, $B=\SI{30}{\milli\tesla}$). The magnetic field is applied along the $y$ direction (vertical axis). The trajectory of the swimming Janus colloid is color-coded from blue to yellow, and the cargo particles are in white to red. The ICEP driving condition is $E=\SI[per-mode=symbol]{0.033}{\volt\per\micro\meter}$ at 2 kHz.\\

\item Movie 3: Reversible control of cross-field motion of active Janus colloidal particles. The applied magnetic field ($B=\SI{25}{\milli\tesla}$) is alternately switched on and off at \SI{60}{\second} intervals. The magnetic field is applied along the $y$ direction (vertical axis).

\end{itemize}

\end{document}